\documentclass[12pt,
showpacs,preprintnumbers,superscriptaddress,nofootinbib]{revtex4}
\usepackage{epsfig}

\begin{document}

\def\simge{\hspace*{0.2em}\raisebox{0.5ex}{$>$}
     \hspace{-0.8em}\raisebox{-0.3em}{$\sim$}\hspace*{0.2em}}
\def\simle{\hspace*{0.2em}\raisebox{0.5ex}{$<$}
     \hspace{-0.8em}\raisebox{-0.3em}{$\sim$}\hspace*{0.2em}}
\def\bra#1{{\langle#1\vert}}
\def\ket#1{{\vert#1\rangle}}
\def\coeff#1#2{{\scriptstyle{#1\over #2}}}
\def\undertext#1{{$\underline{\hbox{#1}}$}}
\def\hcal#1{{\hbox{\cal #1}}}
\def\sst#1{{\scriptscriptstyle #1}}
\def\eexp#1{{\hbox{e}^{#1}}}
\def\rbra#1{{\langle #1 \vert\!\vert}}
\def\rket#1{{\vert\!\vert #1\rangle}}
\def\lsim{{ <\atop\sim}}
\def\gsim{{ >\atop\sim}}
\def\nubar{{\bar\nu}}
\def\psibar{{\bar\psi}}
\def\Gmu{{G_\mu}}
\def\alr{{A_\sst{LR}}}
\def\wpv{{W^\sst{PV}}}
\def\evec{{\vec e}}
\def\notq{{\not\! q}}
\def\notk{{\not\! k}}
\def\notp{{\not\! p}}
\def\notpp{{\not\! p'}}
\def\notder{{\not\! \partial}}
\def\notcder{{\not\!\! D}}
\def\notA{{\not\!\! A}}
\def\notv{{\not\!\! v}}
\def\Jem{{J_\mu^{em}}}
\def\Jana{{J_{\mu 5}^{anapole}}}
\def\nue{{\nu_e}}
\def\mn{{m_\sst{N}}}
\def\mns{{m^2_\sst{N}}}
\def\me{{m_e}}
\def\mes{{m^2_e}}
\def\mq{{m_q}}
\def\mqs{{m_q^2}}
\def\mz{{M_\sst{Z}}}
\def\mzs{{M^2_\sst{Z}}}
\def\mws{{M^2_\sst{W}}}
\def\ubar{{\bar u}}
\def\dbar{{\bar d}}
\def\sbar{{\bar s}}
\def\qbar{{\bar q}}
\def\sstw{{\sin^2\theta_\sst{W}}}
\def\gv{{g_\sst{V}}}
\def\ga{{g_\sst{A}}}
\def\pv{{\vec p}}
\def\pvs{{{\vec p}^{\>2}}}
\def\ppv{{{\vec p}^{\>\prime}}}
\def\ppvs{{{\vec p}^{\>\prime\>2}}}
\def\qv{{\vec q}}
\def\qvs{{{\vec q}^{\>2}}}
\def\xv{{\vec x}}
\def\xpv{{{\vec x}^{\>\prime}}}
\def\yv{{\vec y}}
\def\tauv{{\vec\tau}}
\def\sigv{{\vec\sigma}}
\def\sst#1{{\scriptscriptstyle #1}}
\def\gpnn{{g_{\sst{NN}\pi}}}
\def\grnn{{g_{\sst{NN}\rho}}}
\def\gnnm{{g_\sst{NNM}}}
\def\hnnm{{h_\sst{NNM}}}

\def\xivz{{\xi_\sst{V}^{(0)}}}
\def\xivt{{\xi_\sst{V}^{(3)}}}
\def\xive{{\xi_\sst{V}^{(8)}}}
\def\xiaz{{\xi_\sst{A}^{(0)}}}
\def\xiat{{\xi_\sst{A}^{(3)}}}
\def\xiae{{\xi_\sst{A}^{(8)}}}
\def\xivtez{{\xi_\sst{V}^{T=0}}}
\def\xivteo{{\xi_\sst{V}^{T=1}}}
\def\xiatez{{\xi_\sst{A}^{T=0}}}
\def\xiateo{{\xi_\sst{A}^{T=1}}}
\def\xiva{{\xi_\sst{V,A}}}

\def\rvz{{R_\sst{V}^{(0)}}}
\def\rvt{{R_\sst{V}^{(3)}}}
\def\rve{{R_\sst{V}^{(8)}}}
\def\raz{{R_\sst{A}^{(0)}}}
\def\rat{{R_\sst{A}^{(3)}}}
\def\rae{{R_\sst{A}^{(8)}}}
\def\rvtez{{R_\sst{V}^{T=0}}}
\def\rvteo{{R_\sst{V}^{T=1}}}
\def\ratez{{R_\sst{A}^{T=0}}}
\def\rateo{{R_\sst{A}^{T=1}}}

\def\mro{{m_\rho}}
\def\mks{{m_\sst{K}^2}}
\def\mpi{{m_\pi}}
\def\mpis{{m_\pi^2}}
\def\mom{{m_\omega}}
\def\mphi{{m_\phi}}
\def\Qhat{{\hat Q}}

\def\FOS{{F_1^{(s)}}}
\def\FTS{{F_2^{(s)}}}
\def\GAS{{G_\sst{A}^{(s)}}}
\def\GES{{G_\sst{E}^{(s)}}}
\def\GMS{{G_\sst{M}^{(s)}}}
\def\GATEZ{{G_\sst{A}^{\sst{T}=0}}}
\def\GATEO{{G_\sst{A}^{\sst{T}=1}}}
\def\mdax{{M_\sst{A}}}
\def\mustr{{\mu_s}}
\def\rsstr{{r^2_s}}
\def\rhostr{{\rho_s}}
\def\GEG{{G_\sst{E}^\gamma}}
\def\GEZ{{G_\sst{E}^\sst{Z}}}
\def\GMG{{G_\sst{M}^\gamma}}
\def\GMZ{{G_\sst{M}^\sst{Z}}}
\def\GEn{{G_\sst{E}^n}}
\def\GEp{{G_\sst{E}^p}}
\def\GMn{{G_\sst{M}^n}}
\def\GMp{{G_\sst{M}^p}}
\def\GAp{{G_\sst{A}^p}}
\def\GAn{{G_\sst{A}^n}}
\def\GA{{G_\sst{A}}}
\def\GETEZ{{G_\sst{E}^{\sst{T}=0}}}
\def\GETEO{{G_\sst{E}^{\sst{T}=1}}}
\def\GMTEZ{{G_\sst{M}^{\sst{T}=0}}}
\def\GMTEO{{G_\sst{M}^{\sst{T}=1}}}
\def\lamd{{\lambda_\sst{D}^\sst{V}}}
\def\lamn{{\lambda_n}}
\def\lams{{\lambda_\sst{E}^{(s)}}}
\def\bvz{{\beta_\sst{V}^0}}
\def\bvo{{\beta_\sst{V}^1}}
\def\Gdip{{G_\sst{D}^\sst{V}}}
\def\GdipA{{G_\sst{D}^\sst{A}}}
\def\fks{{F_\sst{K}^{(s)}}}
\def\FIS{{F_i^{(s)}}}
\def\fpi{{F_\pi}}
\def\fk{{F_\sst{K}}}

\def\RAp{{R_\sst{A}^p}}
\def\RAn{{R_\sst{A}^n}}
\def\RVp{{R_\sst{V}^p}}
\def\RVn{{R_\sst{V}^n}}
\def\rva{{R_\sst{V,A}}}
\def\xbb{{x_B}}

\def\PR#1{{{\em   Phys. Rev.} {\bf #1} }}
\def\PRC#1{{{\em   Phys. Rev.} {\bf C#1} }}
\def\PRD#1{{{\em   Phys. Rev.} {\bf D#1} }}
\def\PRL#1{{{\em   Phys. Rev. Lett.} {\bf #1} }}
\def\NPA#1{{{\em   Nucl. Phys.} {\bf A#1} }}
\def\NPB#1{{{\em   Nucl. Phys.} {\bf B#1} }}
\def\AoP#1{{{\em   Ann. of Phys.} {\bf #1} }}
\def\PRp#1{{{\em   Phys. Reports} {\bf #1} }}
\def\PLB#1{{{\em   Phys. Lett.} {\bf B#1} }}
\def\ZPA#1{{{\em   Z. f\"ur Phys.} {\bf A#1} }}
\def\ZPC#1{{{\em   Z. f\"ur Phys.} {\bf C#1} }}
\def\etal{{{\em   et al.}}}

\def\delalr{{{delta\alr\over\alr}}}
\def\pbar{{\bar{p}}}
\def\lamchi{{\Lambda_\chi}}
\newcommand{\amulbl}{a_\mu^{\sst{LL}}}
\def\rnu{{R_\nu}}
\def\rnubar{{R_{\bar\nu}}}
\def\sinhat{\sin^2\hat\theta_W}

\newcommand{\slashq}{\not{\hbox{\kern-3pt $q$}}}
\def\stilde{\widetilde}
\newcommand{\beqa}{\begin{eqnarray}}
\newcommand{\eeqa}{\end{eqnarray}}
\newcommand{\beq}{\begin{equation}}
\newcommand{\eeq}{\end{equation}}



\preprint{CALT-68-2422}
\preprint{MAP-286}
\preprint{hep-ph/0301208}

\title{Supersymmetric Effects in Deep Inelastic Neutrino-Nucleus Scattering}


\author{A. Kurylov}
\affiliation{
California Institute of Technology,
Pasadena, CA 91125\ USA}

\author{M.J. Ramsey-Musolf}
\affiliation{
California Institute of Technology,
Pasadena, CA 91125\ USA}
\affiliation{
Department of Physics, University of Connecticut,
Storrs, CT 06269\ USA}
\affiliation{
Institute for Nuclear Theory, University of Washington,
Seattle, WA
98195\ USA
}

\author{S.~Su}
\affiliation{
California Institute of Technology,
Pasadena, CA 91125\ USA}



\begin{abstract}

We compute the supersymmetric (SUSY) contributions to $\nu$
(${\bar\nu}$)-nucleus deep inelastic
scattering in the Minimal Supersymmetric Standard Model (MSSM). We consider
the ratio of neutral
current to charged current cross sections, $\rnu$ and $\rnubar$, and
compare with the deviations
of these quantities from the Standard Model predictions implied by the
recent NuTeV measurement.
After performing a model-independent analysis,
we find that SUSY loop corrections generally have the opposite sign from
the NuTeV anomaly. We discuss
one scenario in which a right-sign effect arises, and show that it is ruled
out by other precision
data. We also  study for R parity-violating (RPV)
contributions. Although RPV effects could,
in principle, reproduce the NuTeV anomaly, such a possibility is also 
ruled out by other precision electroweak measurements.

\end{abstract}


\maketitle

\vspace{0.3cm}

\pagenumbering{arabic}


\section{Introduction}
\label{intro}

Neutrino  scattering experiments have played a key role in elucidating the
structure of the
Standard Model (SM). Recently, the NuTeV collaboration has performed a
precise determination of the ratio $\rnu$
($\rnubar$) of neutral current (NC) and charged current (CC) deep-inelastic
$\nu_\mu$ ($\bar\nu_\mu$)-nucleus cross sections\cite{nutev02}, which can be 
expressed as:
\begin{equation}
\label{eq:rnudef}
R_{\nu ({\bar\nu})} = (g_L^{\rm eff})^2 + r^{(-1)} (g_R^{\rm eff})^2\ \ \ ,
\end{equation}
where $r=\sigma^{CC}_{{\bar\nu} N}/\sigma^{CC}_{\nu N}$ and
$(g_{L,R}^{\rm eff})^2$ are effective hadronic couplings (defined below).
Comparing
the SM predictions\cite{pdg,erler} for $(g_{L,R}^{\rm eff})^2$ with the values
obtained by the NuTeV
Collaboration yields deviations\footnote{We use the quoted experimental
errors on $\rnu$ and
$\rnubar$, rather than adding the errors on $(g_{L,R}^{\rm eff})^2$ in
quadrature, since the latter are correlated and derived from the experimental
cross section ratios.} $\delta R_{\nu({\bar\nu})}=
R_{\nu({\bar\nu})}^{\rm exp}-R_{\nu({\bar\nu})}^{\rm SM}$
\begin{equation}
\delta R_{\nu}=-0.0033 \pm 0.0007, \ \ \
\delta R_{\bar{\nu}}=-0.0019 \pm 0.0016.
\label{eq:deltaRnu}
\end{equation}

Within the SM, these results may be interpreted as a test of the
scale-dependence of the $\sstw$ since the $(g_{L,R}^{\rm eff})^2$ depend on the
weak mixing angle. While the
SM prediction for $\sstw$ at $\mu=\mz$ has been confirmed with high
precision at LEP and SLC, the predicted running
of this parameter to lower scales has yet to be studied systematically. The
results from the NuTeV measurement
imply a $+3\sigma$ deviation at $\mu\sim 10$ GeV, while the current value of
the cesium weak charge, extracted
from atomic parity-violation (PV), implies agreement with the predicted SM
running at a much lower scale\cite{apv}. Measurements
of the PV asymmetries in polarized $ee$ and $ep$ scattering will provide
further tests of this running at
$\mu\sim 0.2$ GeV \cite{slac,qweak}.

This interpretation of the NuTeV results has been the subject of some
debate. Unaccounted for QCD effects, such as
charge symmetry-breaking in parton distributions or nuclear
shadowing\cite{miller}, have
been proposed as possible remedies for the
anomaly. Alternatively, one may consider physics beyond the SM, as reviewed
in Ref. \cite{davidson}.  In what follows, we focus
on one new physics scenario, namely, supersymmetry (SUSY). While a brief
discussion
of SUSY is given in Ref. \cite{davidson}, an extensive, detailed treatment
has yet to appear in
the literature. Because SUSY is one of the most strongly-motivated
extensions of the SM, undertaking such an analysis is a timely endeavor.
The goal of the
present study is to provide this comprehensive treatment.

In performing our analysis, we work within
the framework of the Minimal Supersymmetric Standard Model (MSSM), which
remains the standard baseline for considering SUSY
effects in precision observables. Typically, MSSM studies adopt one or more
models
of SUSY-breaking mediation, thereby vastly simplifying the task of
analyzing the MSSM
parameter space. Here, however, we carry out a
model-independent treatment, avoiding the choice of a
specific mechanism for SUSY-breaking mediation. Since generic features of
the superpartner spectrum implied by the most widely adopted SUSY-breaking
models
may not be consistent with precision data \cite{kurylov02}, we wish to
determine whether there exist any choices for
MSSM parameters that could account for the NuTeV result, even if such
choices lie outside the purview of standard
SUSY-breaking models.

We find that it is difficult -- if not impossible -- to choose MSSM
parameters so as to improve agreement with
the NuTeV result. When R parity is conserved and SUSY effects only arise
via radiative corrections, the magnitude
of their contribution is generally too small for the NuTeV anomaly to
generate significant constraints.
Moreover, for nearly all parameter choices,
the {\em sign} of the SUSY corrections to $\rnu$ and $\rnubar$ is opposite
to that of the NuTeV
anomaly.  We do find
one scenario -- not considered in Ref. \cite{davidson} -- under which SUSY
loops generate a right sign effect with
relatively large magnitude, but this scenario is presently inconsistent
with other precision, electroweak data.

We also consider possible tree-level contributions from R parity-violating
(RPV) interactions, whose presence would render the lightest supersymmetric
particle (LSP)
unstable. We observe that purely leptonic RPV
effects, which arise via the definition of $\sstw$, could in principle also
generate a
right sign effect to account for the NuTeV anomaly. However, precision data
also limits the magnitude of this contribution to be considerably smaller
than necessary. On the other hand, the NuTeV result does yield new
constraints on possible semileptonic RPV interactions.

In short, our qualitative conclusions agree with those of
Ref. \cite{davidson}, though we believe we have
carried out a more exhaustive analysis. Thus, one would have to look to
more exotic new physics possibilities, such as
a $Z'$ boson coupling only to second generation fermions, if SM QCD effects
are ultimately unable to explain the NuTeV
anomaly. We note that the situation here contrasts with that for the PV
electron scattering asymmetries, where the SM
contribution is fortuitously suppressed and where SUSY radiative
corrections could, in principle, produce observable
contributions\cite{kurylov02b}.

The analysis leading to these conclusions is organized in the remainder of
the paper
as follows. In Section II, we discuss
general features of MSSM contributions to $\nu_\mu$ ($\bar\nu_\mu$)-nucleus
scattering, including both SUSY loop corrections
and RPV effects. Section III gives details of the loop computation as well
as the scan over the MSSM parameter space.
In Section IV, we give the RPV analysis, including results of a fit to
other precision data. Section V contains a
summary of our results. Explicit formulae for loop corrections 
with relevant Feynman rules are given in
the Appendices.


\section{$\nu_\mu N$ (${\bar\nu_\mu} N$) Scattering in the MSSM: General
Considerations}

For momentum transfers $q^\mu$ satisfying $|q^2| << \mzs$, the neutrino-quark
interactions can be represented
with sufficient accuracy by an effective four fermion Lagrangian:
\begin{eqnarray}
{\cal L}_{\nu q}^{NC}  & = & -{G_\mu\rho^{NC}_{\nu N}\over\sqrt{2}}
{\bar\nu}_\mu\gamma^\lambda (1-\gamma_5) \nu_\mu
\sum_q {\bar q}\gamma_\lambda [2\epsilon_L^q P_L+2\epsilon_R^q P_R]q \\
{\cal L}_{\nu q}^{CC} & = & -{G_\mu\rho^{CC}_{\nu N}\over\sqrt{2}}
{\bar\mu}\gamma^\lambda (1-\gamma_5) \nu_\mu
{\bar u}\gamma_\lambda(1-\gamma_5) d + {\rm h.c.} \ \ \ ,
\end{eqnarray}
where $P_L=(1-\gamma_5)/2$, $P_R=(1+\gamma_5)/2$  and
\begin{eqnarray}
\epsilon_L^q & = & I_L^3 - Q_q\kappa_\nu\sstw + \lambda_L^q \\
\epsilon_R^q & = & -Q_q\kappa_\nu\sstw +\lambda_R^q \ \ \ .
\end{eqnarray}
The parameters $\rho_{\nu N}^{NC}=\rho_{\nu N}^{CC}=\kappa_\nu=1$ and
$\lambda_{L,R}^q=0$ at tree-level
in the SM. These quantities differ from their tree-level values when ${\cal
O}(\alpha)$ corrections in
the SM or MSSM are included or when other new physics contributions arise.
The precise values of these
quantities individually are renormalization scheme-dependent. 
When computing MSSM loop
contributions, we use the
modified dimensional reduction (${\overline{DR}}$) scheme, in which the
spatial dimension of momenta
are continued into $d=4-2\epsilon$ dimensions while the Dirac matrices
remain four dimensional, as
required by SUSY invariance. Quantities renormalized in the
${\overline{DR}}$ scheme
will be indicated by a hat. Note
also that for the neutrino reactions of interest here, $\rho_{\nu N}^{NC}$,
$\rho_{\nu N}^{CC}$, and $\kappa_\nu$ are universal
(independent of quark flavor), while the $\lambda_{L,R}^q$ are
flavor-dependent.

The NC to CC cross section ratios $\rnu$ and $\rnubar$ can be expressed in
terms of the above parameters via the effective couplings $(g_{L,R}^{\rm
eff})^2$
appearing in Eq. (\ref{eq:rnudef}) in
a straightforward way:
\begin{equation}
\label{eq:glrdef}
(g_{L,R}^{\rm eff})^2 = \left({{\hat M}_Z^2\over {\hat M}_W^2}\right)^2\left(
{{\hat M}_W^2-q^2\over {\hat M}_Z^2-q^2}\right)^2\left({\rho_{\nu N}^{NC}\over
\rho_{\nu N}^{CC}}\right)^2\sum_q\ (\epsilon_{L,R}^q)^2 \ \ \ .
\end{equation}
The SM values for these quantities are\cite{pdg,erler} $(g_L^{\rm
eff})^2=0.3042$
and $(g_R^{\rm eff})^2 = 0.0301$ while the NuTeV results imply
$(g_L^{\rm eff})^2=0.3005\pm 0.0014$ and $(g_R^{\rm eff})^2=0.0310\pm 0.0011$.

\begin{figure}
\resizebox{12. cm}{!}{\includegraphics*[20,490][480,730]{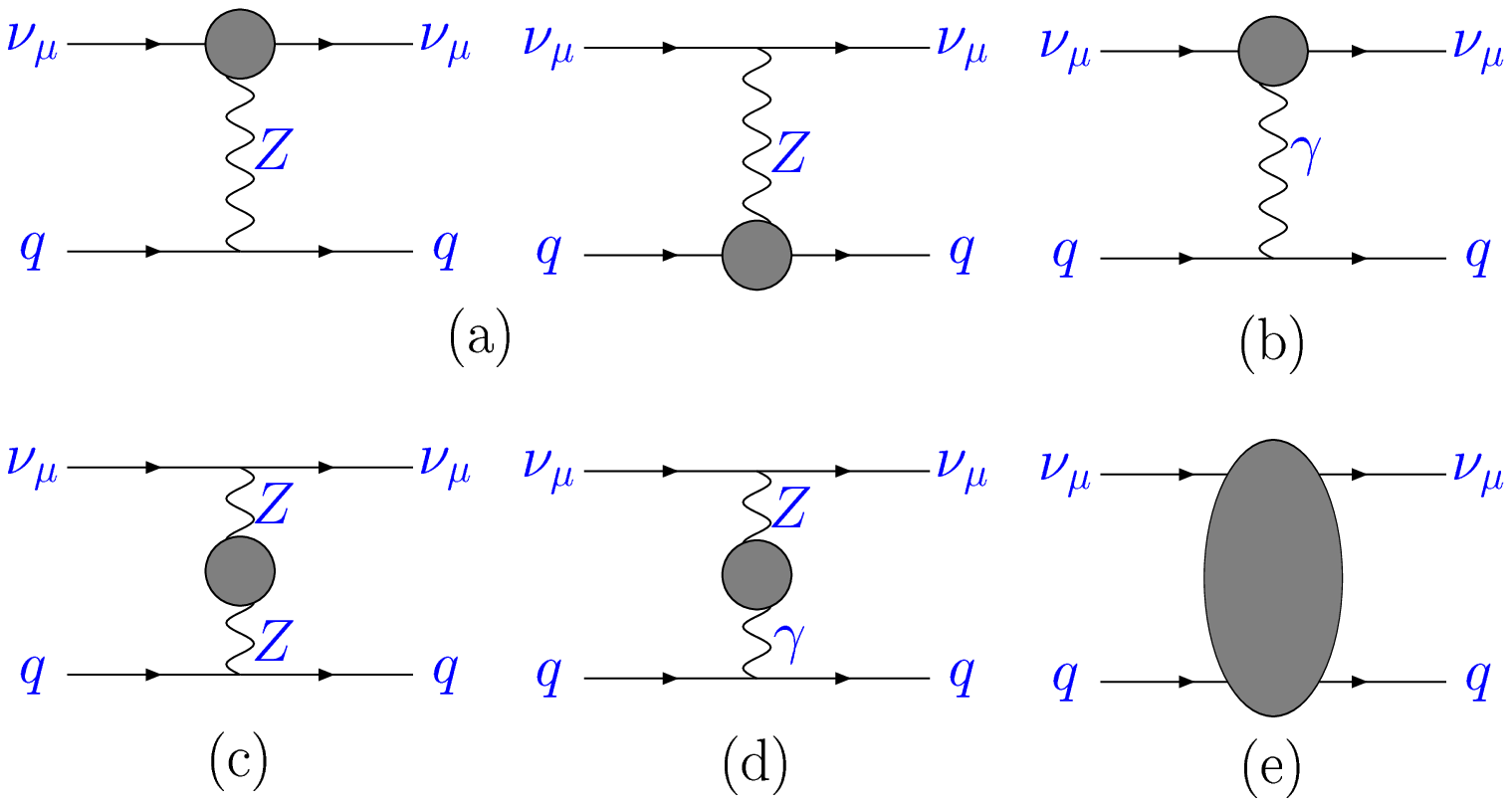}}
\caption{One loop contributions to the neutrino-quark neutral current
amplitude.  The blob
denotes the one loop irreducible diagram or the counter term.  
The Feynman diagrams for the external leg corrections are not 
shown explicitly.}
\label{fig:NC}
\end{figure}

\begin{figure}
\resizebox{16. cm}{!}{\includegraphics*[0,610][600,730]{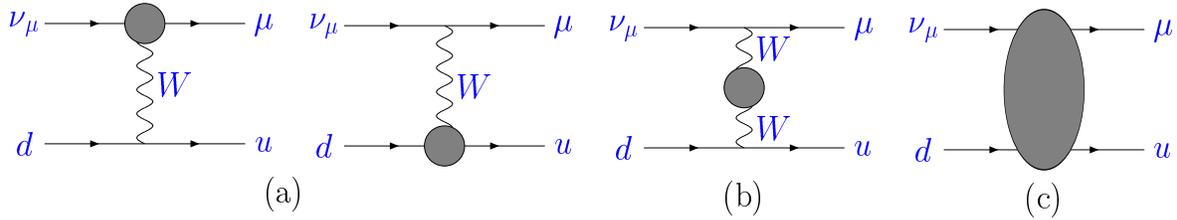}}
\caption{One loop contributions to the neutrino-quark charge current
amplitude.  The blob
denotes the one loop irreducible diagram or the counter term.
The Feynman diagrams for the external leg corrections are not 
shown explicitly.}
\label{fig:CC}
\end{figure}

In what follows, we concentrate on the MSSM contributions to
$\rho^{NC,CC}_{\nu N}$, $\kappa_\nu$,
$\lambda_{L,R}^q$, $\rnu$, and $\rnubar$. When the
R-parity quantum number $(-1)^{3(B-L)+2s}$ is conserved, the MSSM
contributes to these quantities
only via loop effects. The relevant diagrams are shown in Figs.
\ref{fig:NC}-\ref{fig:CC} and the Appendices.
Note that all gauge boson self
energy corrections, as well as leptonic vertex and external
leg corrections,
contribute only to the
universal parameters $\rho_{\nu N}^{NC, CC}$ and $\kappa_\nu$. For the NC
amplitudes, non-universal box diagrams and
quark vertex and external leg corrections (${\hat\delta}_{VB}^{L,R;q}$) are
contained in the
$\lambda_{L,R}^q$. All CC box graphs
as well as hadronic vertex and external leg contributions
(${\hat\delta}_{VB}^{CC}$) appear in
$\rho_{\nu N}^{CC}$. The CC
box graphs also generate amplitudes involving products of scalar and
pseudoscalar currents. However, the
corresponding ${\cal O}(\alpha)$ corrections to $\rnu$ and $\rnubar$ are
suppressed by lepton and quark
masses, so we neglect them here. In terms of these various corrections, the
renormalized parameters are
\begin{eqnarray}
\rho_{\nu N}^{CC} & = & 1 + {{\hat\Pi}_{WW}(q^2)\over
\mws-q^2}-{{\hat\Pi}_{WW}(0)\over \mws} +{\hat\delta}_{VB}^{CC}
-{\hat\delta}_{VB}^{\mu} 
\label{eq:rhoCC}
\\
\rho_{\nu N}^{NC} & = & 1 +{{\hat\Pi}_{ZZ}(q^2)\over\mzs-q^2}
-{{\hat\Pi}_{WW}(0)\over\mws}+{\hat\delta}_V^{\nu}-{\hat\delta}_{VB}^\mu 
\label{eq:rhoNC}
\\
\kappa_\nu & = & {{\hat c}\over{\hat s}}
{{\hat\Pi}_{Z\gamma}(q^2)\over q^2} +{\delta\sinhat\over\sinhat}
 - 4{\hat c}^2 {\hat F}_{A,\nu}(q^2)\\
\lambda_{L,R}^q & = & {\hat\delta}_{VB}^{L,R;q}\ \ \ ,
\label{eq:lambdaNC}
\end{eqnarray}
where ${\hat\Pi}_{VV'}(q^2)$ are the gauge boson self-energies renormalized
in the ${\overline{DR}}$ scheme at a scale $\mu=\mz$;
${\hat\delta}_{VB}^\mu$ denote vertex, external leg, and box graph
corrections entering the muon-decay amplitude; and ${\hat\delta}_V^\nu$
indicates the
neutral current neutrino vertex and lepton external 
leg correction.  Note that 
the muon decay corrections enter the semileptonic amplitudes since the Fermi constant $G_\mu$
is taken from the muon lifetime. 
The correction ${\hat F}_{A,\nu}$ arises from the $\nu_\mu$ charge
radius.\footnote{The $\nu_\mu$ charge radius
is equivalent to its anapole moment in the MSSM.} 
Superpartner loop contributions to ${\hat\Pi}_{Z\gamma}(q^2=0)$ are zero, so
that the corresponding
contributions to $\kappa_\nu$ are finite at the photon point. The shift
in $\sinhat$ arises from its definition in terms of $\alpha$, $G_\mu$, and
$\mz$
\begin{equation}
\label{eq:sinhatdef}
{\hat s}^2{\hat c}^2={\pi\alpha\over\sqrt{2} M_Z^2 G_\mu[1-\Delta {\hat
r}(M_Z)]}
\end{equation}
${\hat s}^2 = 1-{\hat c}^2 =\sinhat(M_Z)$ and\cite{pierce}
\begin{equation}
\Delta{\hat r} = {{\hat\Pi}^{\prime}}_{\gamma\gamma}(0)+2{{\hat s}\over
{\hat c}}
{{\hat \Pi}_{Z \gamma }(0)\over M_Z^2}-{{\hat \Pi}_{ZZ}(M_Z^2)\over M_Z^2}
+{{\hat \Pi}_{WW}(0)\over M_W^2}+{\hat\delta}_{VB}^\mu
\end{equation}
with ${{\hat\Pi}^{\prime}}_{\gamma\gamma}(q^2) 
= {\hat\Pi}_{\gamma \gamma}(q^2)/q^2$.
Writing $\Delta{\hat r} =\Delta{\hat r}^{\rm SM} + \Delta{\hat r}^{\rm
SUSY}$ one
has
\begin{equation}
{\delta{\hat s}^2_{\rm SUSY} \over{\hat s}^2} = {{\hat c}^2\over{\hat
c}^2-{\hat
s}^2} \Delta{\hat r}^{\rm SUSY}\ \ \ .
\end{equation}
In Section III, we discuss our computation of the MSSM loop contributions
to these parameters in detail; explicit expression for the various loop
amplitudes appears in the Appendices.

\begin{figure}
\resizebox{12. cm}{!}{\includegraphics*[10,520][480,740]{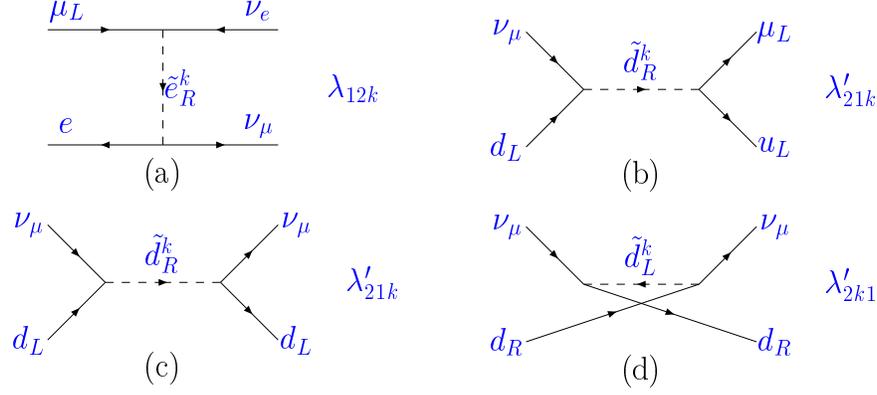}}
\caption{Tree level RPV contributions to muon decay [plot (a)], $\nu_\mu-q$
charged current [plot (b)] and $\nu_\mu-q$ neutral current [plots (c) and (d)]
amplitudes. }
\label{fig:RPV}
\end{figure}

When $R$-parity is not conserved, however, new tree-level contributions appear.
The latter are generated by the $B-L$ violating
superpotential
\begin{equation}
\label{eq:rpvsuper}
W_\sst{RPV} = \frac{1}{2}\lambda_{ijk}L_i L_j {\bar E}_k
+\lambda^{\prime}_{ijk}L_i Q_j{\bar D}_k +
\frac{1}{2}\lambda^{\prime\prime}_{ijk}{\bar U}_i{\bar D}_j{\bar D}_k\ \ \ ,
\end{equation}
where $L_i$ and $Q_i$ denote lepton and quark SU(2)$_L$ doublet
superfields, $E_i$, $U_i$, and $D_i$ are
singlet superfields and the $\lambda_{ijk}$ {\em etc.} are {\em a priori}
unknown couplings. 
We have neglected additional lepton-Higgs mixing 
term in Eq.~(\ref{eq:rpvsuper}) for simplicity.
In order to
avoid unacceptably large contributions to the proton decay rate, we set the
$\Delta B\not= 0$ couplings
$\lambda^{\prime\prime}_{ijk}$ to zero. The purely leptonic terms
($\lambda_{12k}$) contribute to neutrino scattering
amplitudes via the normalization of CC and NC amplitudes to $G_\mu$ and
through the definition of
$\sinhat$\cite{mrm00}.  The remaining semileptonic, $\Delta L=\pm 1$
interactions ($\lambda^{\prime}_{ijk}$) give
direct contributions to the neutrino scattering amplitudes. The latter may be
obtained computing the Feynman amplitudes in Fig. \ref{fig:RPV}
and performing a Fierz reordering. For neutrino-quark scattering, one
obtains the effective Lagrangian
\begin{eqnarray}
\label{eq:rpveffective}
{\cal L}_\sst{RPV}^\sst{EFF} & = & -{|\lambda^{\prime}_{2k1}|^2\over 2
M^2_{\tilde d^k_L}}{\bar d}_R\gamma^\mu d_R
{\bar\nu}_{\mu L}\gamma_\mu \nu_{\mu L} +
{|\lambda^{\prime}_{21k}|^2\over 2 M^2_{\tilde d^k_R}}{\bar d}_L\gamma^\mu
d_L {\bar\nu}_{\mu L}\gamma_\mu \nu_{\mu L}\nonumber \\
&&  - {|\lambda^{\prime}_{21k}|^2\over 2 M^2_{\tilde d^k_R}}\Biggl[{\bar
u}_L\gamma^\mu d_L
{\bar\mu}_{ L}\gamma_\mu \nu_{\mu L}+{\rm h.c.}\Biggr]\ \ \ ,
\end{eqnarray}
where we have taken $|q^2| << M_{\tilde f}^2$ and have retained only the
semileptonic terms relevant to $\nu_\mu$-$q$ scattering.

In terms of these parameters, the shifts induced in the effective $\nu_\mu$
(${\bar\nu_\mu}$)-$N$ parameters
$\rho_{\nu N}^{CC,\ NC}$ and $\epsilon^q_{L,R}$ are
\begin{eqnarray}
\delta\rho_{\nu N}^{NC} & = & \Delta_{12k}({\tilde e}^k_R) \\
\delta\rho_{\nu N}^{CC} & = & \delta\rho_{\nu N}^{NC} +
\Delta^{\prime}_{21k}({\tilde d}^k_R) \\
\delta\epsilon_L^d & = & -\Delta^{\prime}_{21k}({\tilde
d}^k_R)+\frac{1}{3}{\lambda_x}
\Delta_{12k}({\tilde e}^k_R)\\
\delta\epsilon_R^d & = & -\Delta^{\prime}_{2k1}({\tilde
d}^k_L)+\frac{1}{3}{\lambda_x}
\Delta_{12k}({\tilde e}^k_R)\\
\delta\epsilon^u_L = \delta\epsilon^u_R & = &
-\frac{2}{3}{\lambda_x}\Delta_{12k}({\tilde e}^k_R)\ \ \ ,
\end{eqnarray}
where
\begin{equation}
\Delta_{ijk}({\tilde f}) = {|\lambda_{ijk}|^2\over 4\sqrt{2} G_\mu
M^2_{\tilde f}}
\end{equation}
and\cite{mrm00}
\begin{equation}
{\lambda_x}\approx {{\hat s}^2 {\hat c}^2\over {\hat c}^2-{\hat
s}^2}\approx 0.35\ \ \ .
\end{equation}
The corresponding shifts in $R_{\nu(\bar\nu)}$ are
\begin{eqnarray}
\label{eq:rnurpv}
\delta R_{\nu (\bar\nu)}&=&\lambda_x
[-\frac{4}{3} \epsilon_L^u + \frac{2}{3}\epsilon_L^d ]
[1+r^{(-1)} ]\Delta_{12k}(\tilde{e}_R^k)
-2[R_{\nu (\bar\nu)}^{\rm SM}+\epsilon_L^d]
\Delta_{21k}^{\prime}(\tilde{d}_{R}^k)\\
\nonumber
&& \ \ \ +2r^{(-1)}\epsilon_R^d
\Delta_{2k1}^{\prime}(\tilde{d}_{L}^k)\\
\nonumber
&\approx& -0.25 [1+r^{(-1)} ]\Delta_{12k}(\tilde{e}_R^k) -
2[R_{\nu (\bar\nu)}^{\rm SM}-0.43]
\Delta_{21k}^{\prime}(\tilde{d}_{R}^k)+ 1.6 r^{(-1)}
\Delta_{2k1}^{\prime}(\tilde{d}_{L}^k)\ \ \ .
\end{eqnarray}

As we discuss in Section IV, $\Delta_{12k}({\tilde e}^k_R)$ and
$\Delta^{\prime}_{21k}({\tilde d}^k_R)$ are
constrained by other precision electroweak data, while
$\Delta^{\prime}_{2k1}({\tilde d}^k_L)$ is relatively
unconstrained. In Eq. (\ref{eq:rnurpv}), the coefficients of
$\Delta^{\prime}_{21k}({\tilde
d}^k_R)$ and $\Delta^{\prime}_{2k1}({\tilde d}^k_L)$ are positive, while
the coefficient
of $\Delta_{12k}({\tilde e}^k_R)$ is negative. Since the $\Delta_{ijk}$ are
non-negative,
we would require sizable value of $\Delta_{12k}({\tilde e}^k_R)$ and rather
small values of $\Delta^{\prime}_{21k}({\tilde d}^k_R)$ and
$\Delta^{\prime}_{2k1}({\tilde d}^k_L)$ 
to account for the negative shifts in $\rnu$ and $\rnubar$ implied
by the
NuTeV analysis. The present constraints on $\Delta_{12k}({\tilde e}^k_R)$,
however, are
fairly stringent, ruling out sizable values for the semileptonic
corrections with fairly
high confidence.

\section{SUSY Loop Contributions}
\label{sec:MSSM}
Instead of working in a specific SUSY breaking scenario, we  perform a
model-independent analysis by varying all the possible soft SUSY-breaking
parameters\cite{haber}.  Although such an approach is insensitive to the
effects of
any particular SUSY breaking parameter, it does allow us to obtain the size of
SUSY contributions in the most general way.  In our analysis, we set 
the momentum transfer $q^2=0$.

We first compute loop effects only (R-parity being conserved)
by scanning over the MSSM parameters in the
ranges shown in Table.~\ref{table:scanparams}. Here,  $\tan\beta=v_u/v_d$
is the
ratio of the up and down type Higgs vacuum expectation values;  $\mu$ is
bilinear Higgs coupling in the supersymmetric Lagrangian, which
gives the mass to the Higgsino;  $M_1$, $M_2$
and $M_{\tilde{g}}$ are the masses for ${\rm U}(1)_Y$, ${\rm SU}(2)_L$ and
${\rm SU}(3)_c$ gaugino, respectively;
$M^i_{\tilde{f}_{L,R}}$ are the diagonal mass parameters
for the left- and right- handed squarks and sleptons of generation $i$, while
$M^i_{\tilde{f}_{LR}}$ are the left-right
mixing parameters. In order to avoid unacceptably large flavor-changing
neutral currents,
we do not allow for flavor mixing between squark generations but
do allow for superpartners of different generations to have different
masses.

\begin{table}
\begin{tabular}{|c|cc|}
\hline
Parameter & Min & Max \\
\hline
$\tan\beta$ & 1.4 & 60 \\
${\tilde M}$ & 50 GeV & 1000 GeV \\
$M^i_{\tilde{f}_{LR}}$
& -1000 GeV & 1000 GeV \\
\hline
\end{tabular}
\caption{Ranges of SUSY parameters scanned.  Here, ${\tilde M}$ denotes
any of $|\mu|$, $M_{1,2,{\tilde g}}$, or
the diagonal sfermion mass parameters
$M^i_{{\tilde f}_{L,R}}$. The $\mu$ parameter can
take either sign.  The generation index $i$ runs from 1 to 3.}
\label{table:scanparams}
\end{table}

In randomly choosing values for these parameters, we follow the conventional
practice of using  a linear
distribution for all parameters except $\tan\beta$, for which we use a
logarithmic
distribution.  We discard any points that yield SUSY particle
masses below present
collider lower bounds.  In addition, we impose  constraints from the
$Z$-pole electroweak precision measurements,
which are embodied in terms of the
three oblique parameters $S$, $T$ and $U$ \cite{peskin}.  
To that end, we
express
the SUSY shifts in $\rho_{\nu N}^{NC} $ and $\kappa_\nu$ in terms of these
quantities:
\begin{eqnarray}
\label{eq:rho-kappa-stu}
\delta\rho^{NC}_{\nu N} & = & {\hat\alpha} T-{\hat\delta}_{VB}^\mu \nonumber
\\
\delta\kappa_{\nu}& = &
\left({{\hat c}^2\over {\hat c}^2-{\hat s}^2} \right)
\left({{\hat\alpha}\over 4{\hat s}^2 {\hat c}^2} S-{\hat \alpha} T
+{\hat\delta}_{VB}^\mu \right) \nonumber \\
&&+{{\hat c}\over {\hat s}}\Bigl[  {{\hat\Pi}_{Z\gamma}(q^2)\over q^2}-
{{\hat\Pi}_{Z \gamma }(M_Z^2)\over M_Z^2}\Bigr]^{\rm SUSY} \nonumber \\
&&+\Bigl({{\hat c}^2\over {\hat c}^2-{\hat s}^2}
\Bigr)\Bigl[-{{\hat\Pi}_{\gamma\gamma}(M_Z^2)\over M_Z^2}
+{\Delta{\hat\alpha}\over{\hat\alpha}}  
\Bigr]^{\rm SUSY}- 4 {\hat c}^2 \hat{F}^{\rm SUSY}_{A,\nu}(q^2)\ \ \ ,
\end{eqnarray}
where remaining terms proportional to $q^2$ are negligible and have been
dropped.
Here $\Delta{\hat\alpha}$ is the SUSY contribution to the 
difference between the
fine structure constant and the electromagnetic coupling renormalized 
at $\mu=M_Z$:
$\Delta {\hat \alpha}=\left[{\hat \alpha}(M_Z)-\alpha\right]^{\rm SUSY}$.

Note that only $S$ and $T$ enter these expressions. Since these parameters
are correlated, we use the 95$\%$ C.L. $S-T$ constraints \cite{pdg} and retain
only those
parameter choices consistent with these constraints. We observe that this
procedure is not entirely self-consistent, since we have not taken into
account non-oblique
corrections to $Z$-pole observables in deriving the oblique parameter
constraints. Nevertheless,
the essential, qualitative implications of precision $Z$-pole data for SUSY
loop effects on the $\nu_\mu$-$q$  parameters are unaffected by this 
inconsistency.
We also omit parameter choices generating too large a SUSY contribution to the
muon anomalous  magnetic moment\cite{muon}. Doing so limits left-right mixing
for second generation sleptons to be fairly small.

Before presenting our results, we also comment on the inputs used in
Ref. \cite{davidson}. In that study,
the authors only investigated the slepton contribution with four MSSM
parameters:
$M_1$, $\mu$, $\tan\beta$ and $M_{\tilde{l}}$, assuming the
GUT relation for the gaugino masses and the slepton mass degeneracy.
In our  analysis, we include the contribution from both the squark and
slepton sectors,
taking into account sfermion left-right mixing,
and allowing for non-universality between generations.

\begin{figure*}
\resizebox{8cm}{!}{
\includegraphics{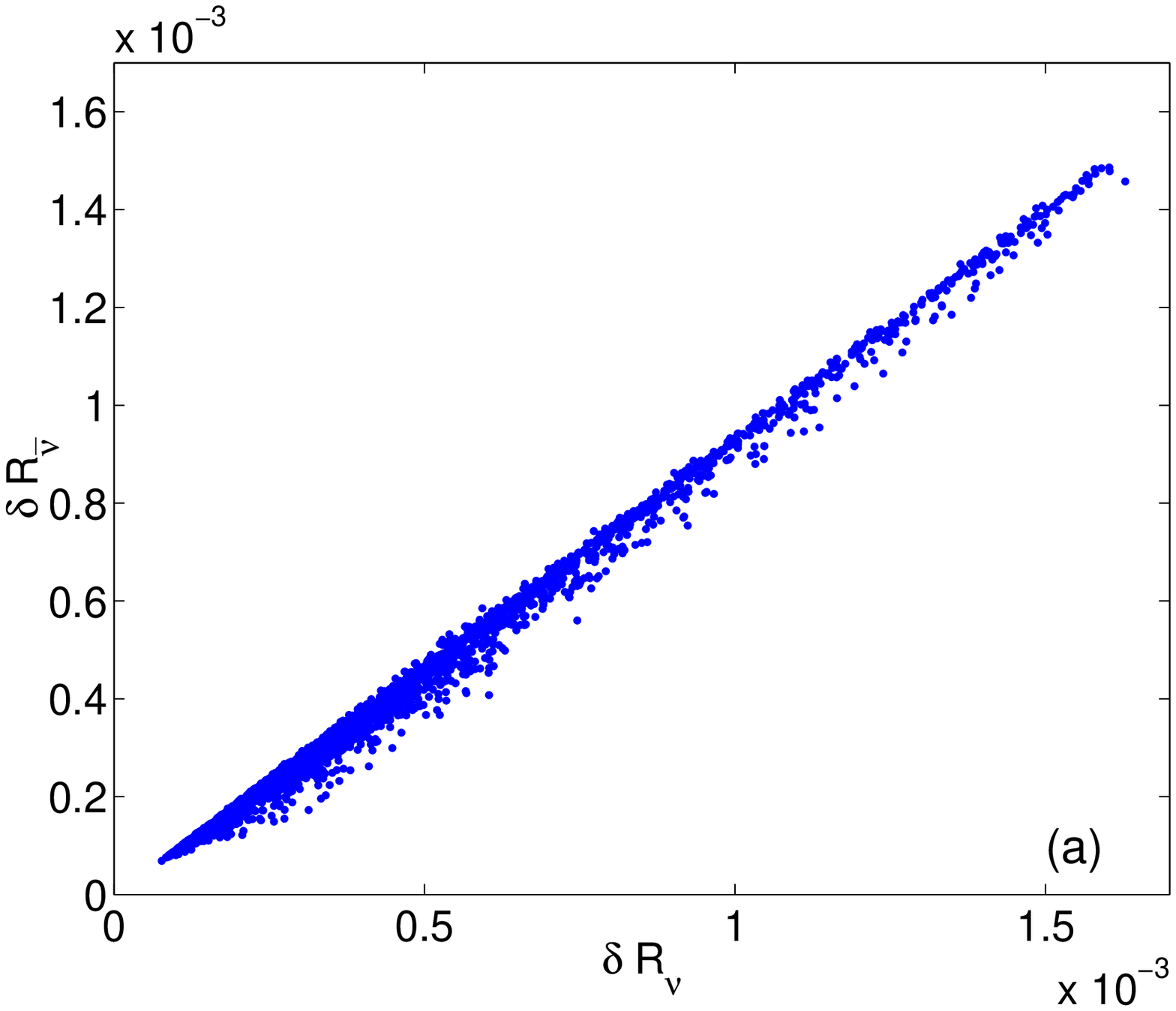}}
\resizebox{8cm}{!}{
\includegraphics{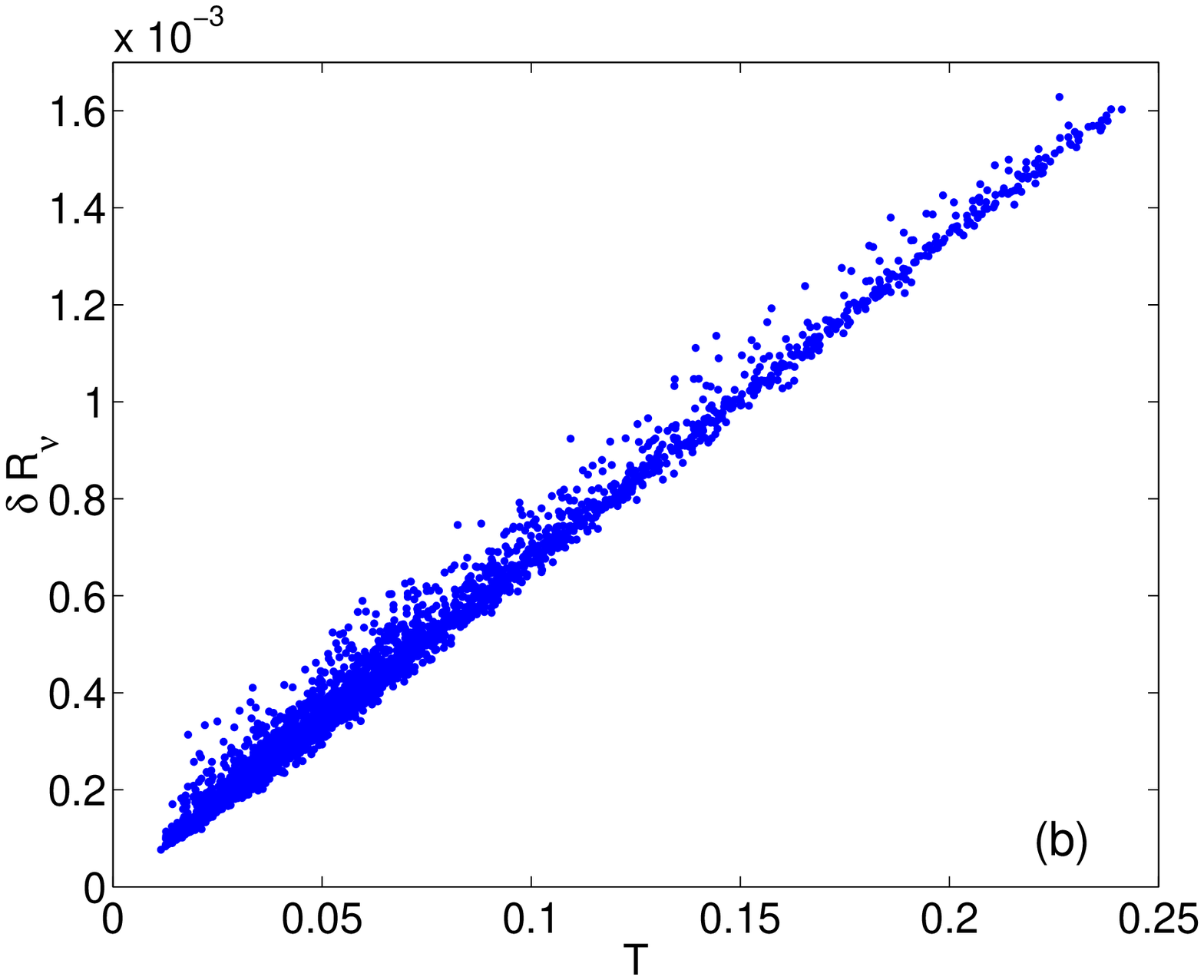}}
\caption{Plot (a) gives the
MSSM contribution to $R^{\nu}$ and $R^{\bar\nu}$ with MSSM parameters 
chosen randomly from 
range shown in Table.~\ref{table:scanparams}.
Plot (b) shows the dependence of $\delta R_{\nu}$ on the oblique
parameter $T$ with the random MSSM parameter set.
}
\label{fig:Rnurandom}
\end{figure*}

\begin{figure}
\resizebox{8cm}{!}{
\includegraphics{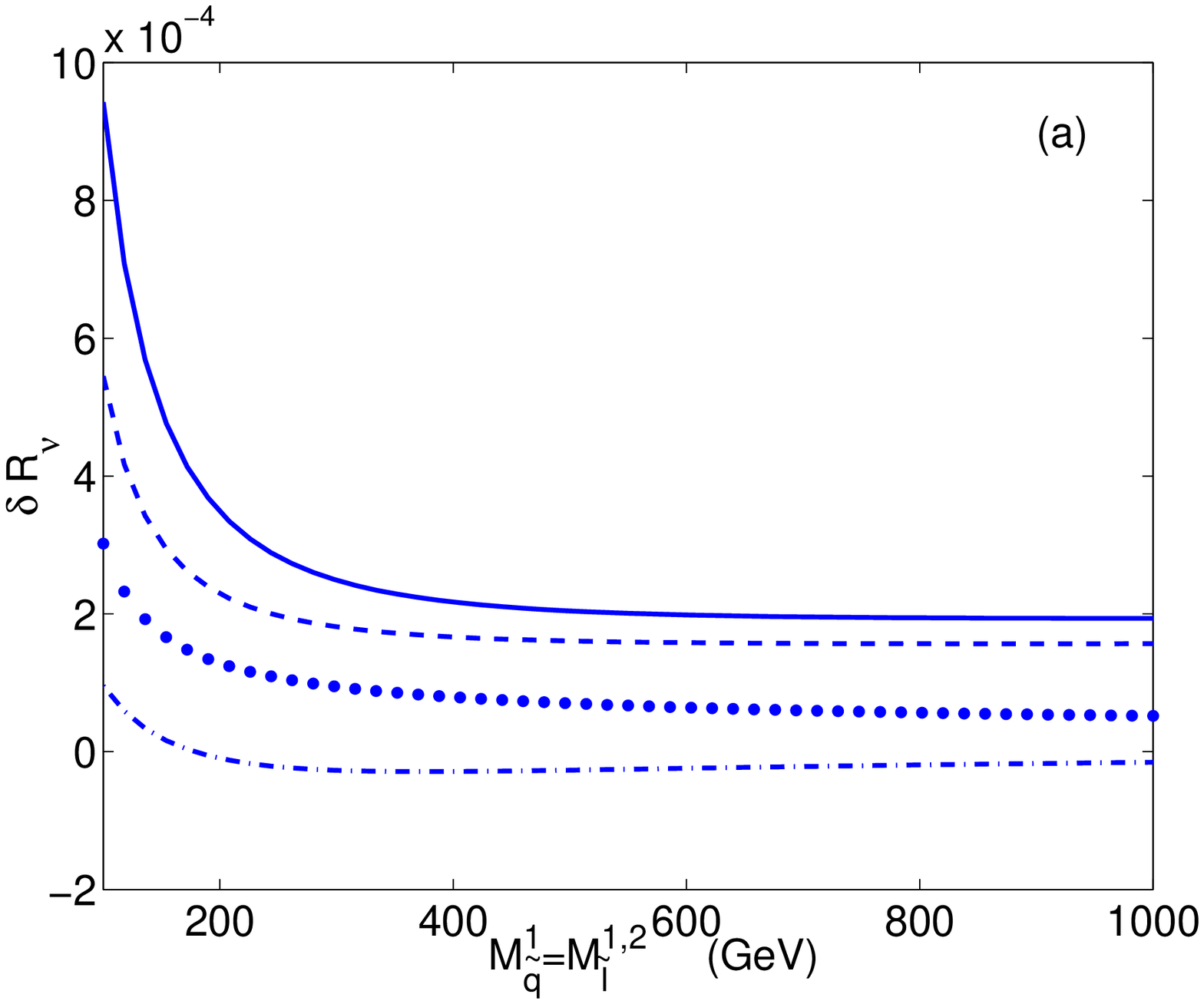}}
\resizebox{8cm}{!}{
\includegraphics{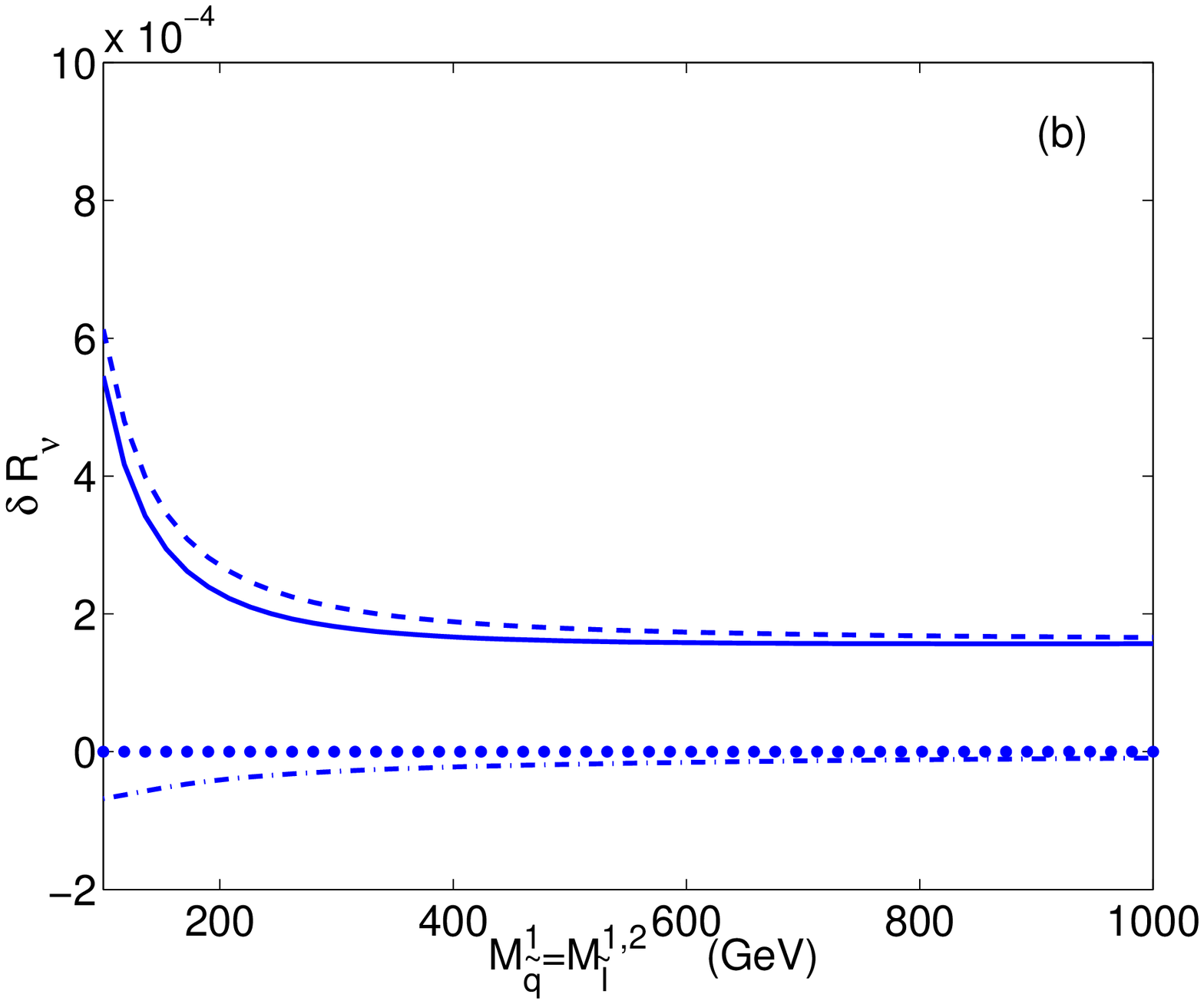}}
\caption{Contributions to $\delta R_{\nu}$ from
$\delta \rho^{NC}-\delta \rho^{CC}$ (dashed line), $\delta \kappa$
(dotted line) and $\lambda^{NC}$ (dash-dotted line).  The solid line is the
sum of all the contributions to $\delta R_{\nu}$.  Plot(b) shows the size
of various components of $\delta \rho^{NC}-\delta \rho^{CC}$ (solid line):
$T$ (dashed line), $\hat\delta_{VB}^{\mu}$ (dotted line) and
$\hat\delta_{VB}^{NC}-\hat\delta_{VB}^{CC}$ (dash-dotted line).  
The $x$-axis is
the common first generation squark mass and first and second generation
slepton mass.  The other MSSM parameters are chosen to be
$\tan\beta=10$, $2M_1=M_2=\mu=200$ GeV.  The mass for the second and third
generation squarks and third generation slepton are taken to be 1000 GeV.
}
\label{fig:Rnu_sep}
\end{figure}

With a sample of about 3000 randomly selected parameter sets, we
calculated the MSSM
contributions to $R_{\nu}$ and $R_{\bar\nu}$.  The numerical results are  shown
in Fig.~\ref{fig:Rnurandom}(a). We observe that $R_{\nu}$
and $R_{\bar\nu}$ are highly correlated. This correlation arises because
MSSM contributions to the $r^{(-1)}(g_R^{\rm eff})^2$ term in
Eq.~(\ref{eq:rnudef}) are
small, while the contribution to the $(g_L^{\rm eff})^2$ terms are the same
for
both $R_{\nu}$ and $R_{\bar\nu}$.  We also find that the magnitude of
$\delta R_{\nu,\bar\nu}$
is dominated by the SUSY contributions to $T$ (see Figs. \ref{fig:Rnurandom}(b)
and \ref{fig:Rnu_sep}), which
is sensitive to
isospin-breaking in the SUSY spectrum.
It is bounded above by other precision electroweak data, thereby limiting the
size of possible
SUSY loop contributions to $\rnu$ and $\rnubar$ to be considerably smaller
than the deviations
in Eq.~(\ref{eq:deltaRnu}). 
A breakdown of the various classes contributions is given in 
Fig.~\ref{fig:Rnu_sep}.
More significantly, the sign of the SUSY loop
corrections is
nearly always positive, in contrast to the sign of the NuTeV anomaly.

\begin{figure*}
\resizebox{8cm}{!}{\includegraphics{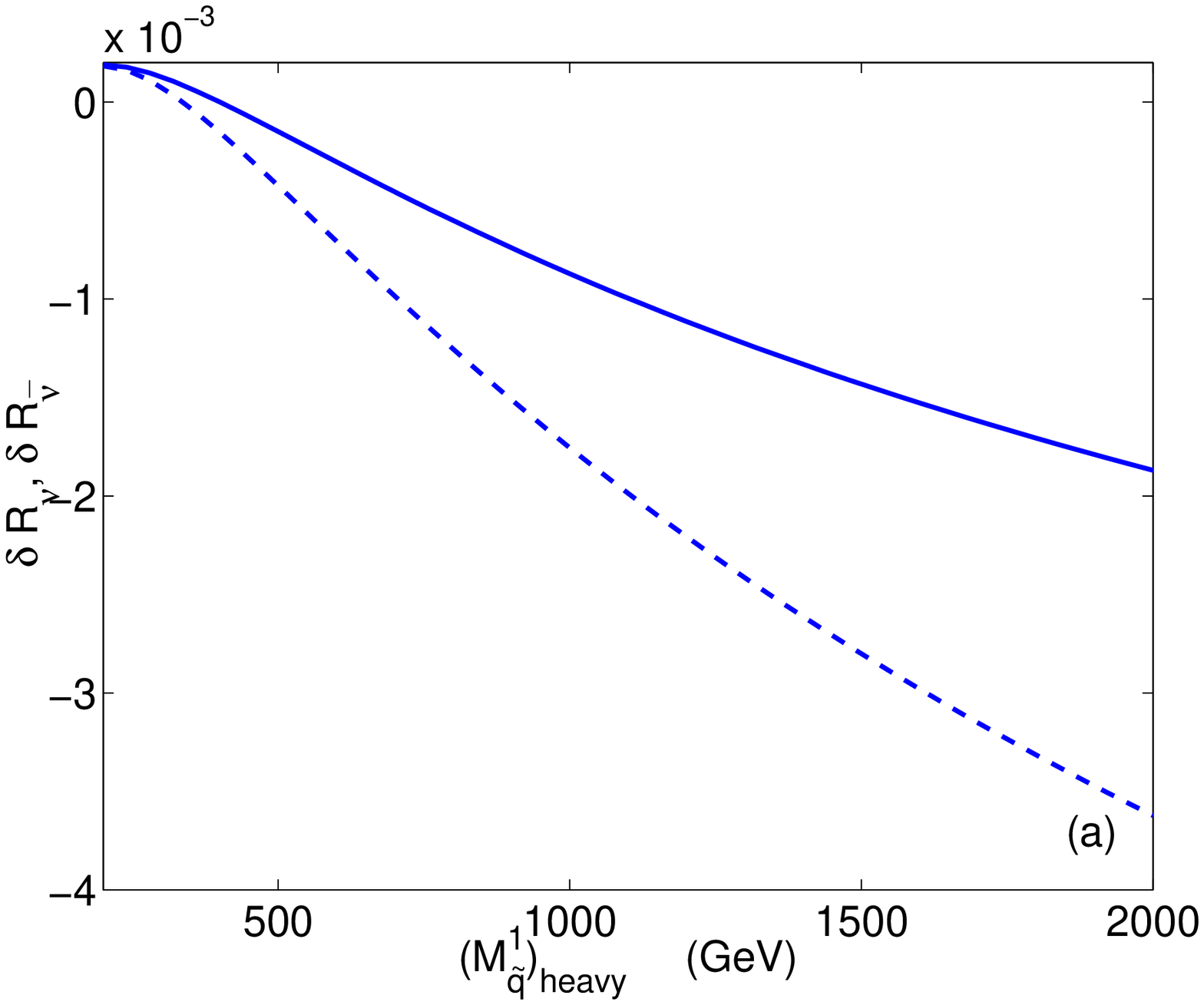}}
\resizebox{8cm}{!}{\includegraphics{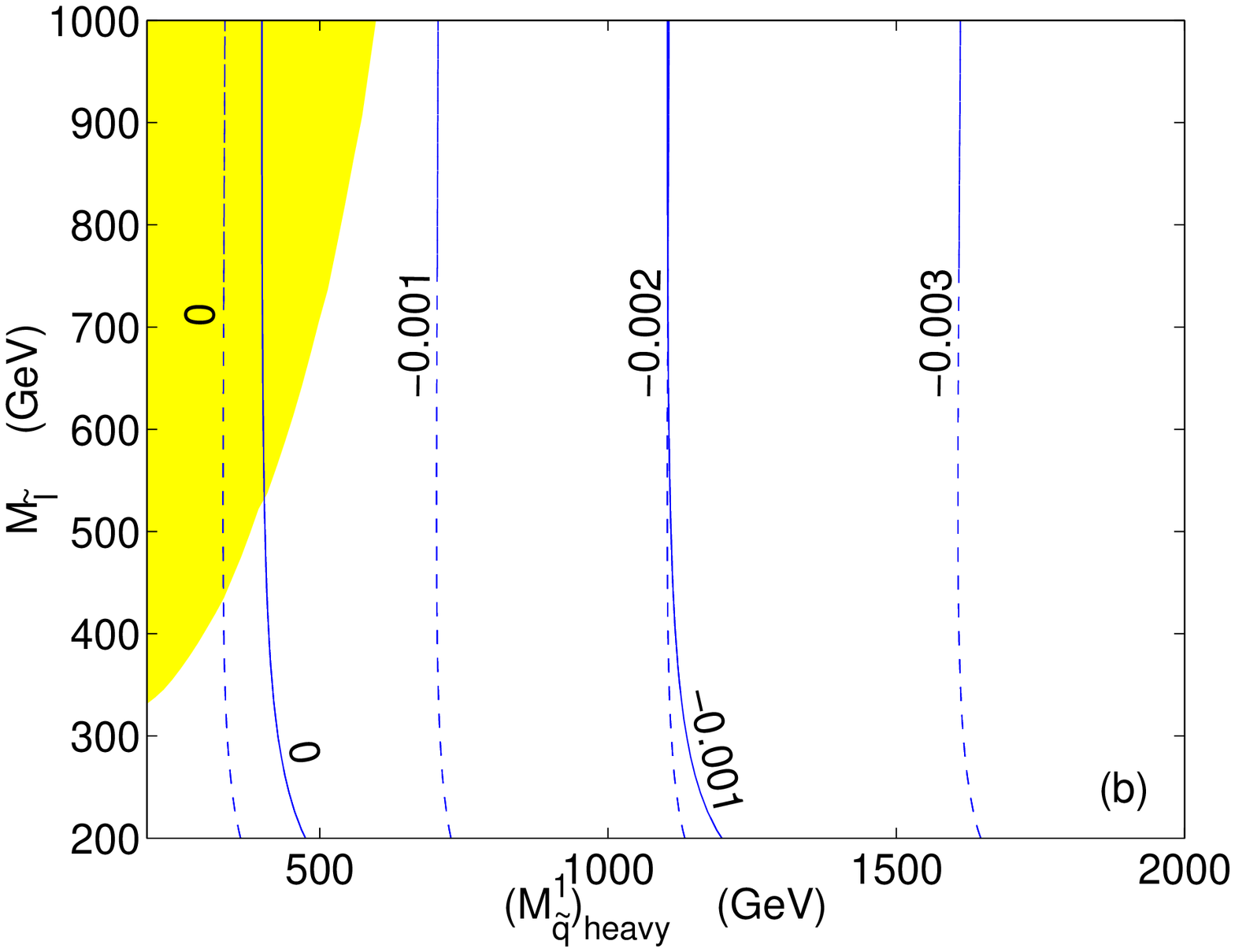}}
\caption{Plot (a) shows the MSSM contribution to $R^{\nu}$ (solid line) and
$R^{\bar\nu}$ (dashed line) with respect to the heavy first
generation squark masses $M_{\tilde{u}}^1=M_{\tilde{d}}^1\equiv
M_{\tilde{q}}^1$,
when the light first generation squark mass is fixed to be 200 GeV and
the left-right mixing angle in fixed to be $\theta_{\tilde{u}}^1=
\theta_{\tilde{d}}^1=\pi/4$.  No $L-R$ mixing for other squarks and sleptons
is assumed. 
The rest of the MSSM
parameters are chosen to be: $2M_1=M_2=\mu=200$ GeV, $M_{\tilde{g}}=200$ GeV,
$M_{\tilde{l}}=1000$ GeV, $M_{\tilde{q}}^{2,3}=1000$ GeV and $\tan\beta=10$.
Plot (b) shows the contours for the shifts $\delta R_{\nu}$ (solid line) and
$\delta R_{\bar\nu}$ (dashed line) in the plane of slepton mass and
heavy first generation squark masses.  Slepton mass degeneracy is assumed.
The rest of the MSSM parameters are the same as those used in plot (a).
The shaded region in (b) shows the region preferred by charge current
universality.
}
\label{fig:gluino}
\end{figure*}

We do, however, find one corner of the parameter space which admits a negative
loop contribution. This scenario involves gluino loops, whose effect can become
large and negative when  the first generation
up-type squark and down-type squarks are nearly degenerate  and
left-right mixing is close to maximal.
In this particular region of the MSSM parameter space, there is no gluino loop
correction
to the quark  charged and neutral vector currents,
while the axial
currents receive large corrections. When $M^{2,3}_{{\tilde f}_{LR}}=0$ and
and the second and third generation soft parameters are sufficiently heavy
(corresponding to a small value for $T$),  the gluino contribution could
give rise to a negative correction to $R_{\nu, \bar\nu}$,
as shown in Fig.~\ref{fig:gluino} (a). In fact, the gluino
contribution could be as much as few$\times 10^{-3}$ in magnitude.  
As illustrated in Fig.
\ref{fig:gluino}(b), however,
equal and large left-right mixing for both up- and down-type squarks
is inconsistent with the other precision electroweak
inputs, such as the $M_W$ and charged current universality\cite{kurylov02}.
The shaded region
in  Fig.~\ref{fig:gluino}(b) shows the region preferred
by these other inputs in the $(M_{\tilde{q}}^1)_{\rm heavy}$
and $M_{\tilde{\ell}}$ plane. Note that a shift in $|V_{us}|$, as implied
by the
recent analysis of charged $K_{\ell 3}$ decays in the E865 experiment at
Brookhaven
National Laboratory\cite{sher}, would increase the shaded region in
Fig.~\ref{fig:gluino}.

It is interesting to note that even if other precision data did not rule
out this gluino loop effect,
it could not account for the apparent deviation of $\sstw$ from the SM
prediction implied by the NuTeV
analysis. The latter relies on a modified version of the
Paschos-Wolfenstein relation\cite{pwrel}
\begin{equation}
R^{-} \equiv {\rnu - r \rnubar\over 1-r} = \frac{1}{2}(1-2\sstw)+ \cdots
\end{equation}
where we have shown the SM prediction for $R^{-}$ with the $+\cdots$
indicating higher-order
corrections. In practice, the extraction of $\sstw$
relies on a slightly different combination of $\rnu$ and $\rnubar$ in the
numerator,
\begin{equation}
{\tilde R}^- \equiv {\rnu-\xi\rnubar\over 1-r}
\end{equation}
with $\xi$ chosen to be slightly different from $r$ in order to minimize
charm quark
mass uncertainties in the $\sstw$ extraction\cite{heidi}. It is
straightforward to show that gluino
loop contributions to ${\tilde R}^-$ are proportional to $\xi-r$, thereby
minimizing
their impact on $\sstw$. Specifically, we find that for maximal left-right 
mixing $\theta_{\tilde{u}}^1=\theta_{\tilde{d}}^1=\pi/4$,
\begin{eqnarray}
{\tilde R}^-&=&{\alpha_s\over 3\pi}(\xi-r){1+r\over
r(1-r)}{\hat s}^2\left(1-{5\over 9}{\hat s}^2\right)\Biggl(
2V_2\left[M_{\tilde g},(M_{\tilde{q}}^{1})_{\rm
heavy},(M_{\tilde{q}}^{1})_{\rm light}\right] \nonumber \\
&-&V_2\left[M_{\tilde g},(M_{\tilde{q}}^{1})_{\rm
heavy},(M_{\tilde{q}}^{1})_{\rm heavy}\right]
-V_2\left[M_{\tilde g},(M_{\tilde{q}}^{1})_{\rm
light},(M_{\tilde{q}}^{1})_{\rm light}\right]
\Biggr)
\end{eqnarray}
where $\alpha_s$ is the strong coupling constant and $V_2(M,m_1,m_2)$ is
defined in Eq.
(\ref{eq:vert-functions}).

In summary, it is difficult to explain the NuTeV anomaly in
the $R$-parity conserving MSSM framework.
In most of the MSSM parameter space, the MSSM contributions to
$R_{\nu}$ and $R_{\bar\nu}$  are small and have the wrong sign.
Gluino loops could give sizable, negative contributions to
$R_{\nu}$ and $R_{\bar\nu}$, when the left-right mixing in the first
generation squark sector is equal and close to maximum.  However, this
scenario, which is inconsistent with charged current data, could not
in any case  account for
the deviation in $\sstw$ implied by the NuTeV analysis.

\section{RPV contributions}
\label{sec:RPV}

As in the case of SUSY loop corrections, the effects of RPV contributions to
$\rnu$ and $\rnubar$ are correlated with similar effects on other precision
electroweak observables\cite{mrm00}. The relevant correlations are
indicated in
Table.~\ref{table:RPV}, where we list the RPV contribution to four
relevant precision observables: superallowed  nuclear $\beta$-decay that
constrains $|V_{ud}|$ \cite{towner}, atomic PV measurements of
the cesium weak charge $Q_W^{Cs}$ \cite{apv}, the $e/\mu$ ratio $R_{e/\mu}$ in
ratio $\pi_{l2}$ decays\cite{pil2}, and a comparison of the
Fermi constant $G_\mu$ with the appropriate combination of
$\alpha$, $M_Z$, and $\sstw$\cite{marciano}. The values of the experimental
constraints on those
quantities are given  in the last column.

\begin{table}
\begin{tabular}{|c|ccccc|r|}
\hline
& $\Delta_{11k}^{\prime}(\tilde{d}_R^k)$
& $\Delta_{1k1}^{\prime}(\tilde{q}_L^k)$
& $\Delta_{12k}(\tilde{e}_R^k)$
& $\Delta_{21k}^{\prime}(\tilde{d}_R^k)$
& $\Delta_{2k1}^{\prime}(\tilde{d}_L^k)$
&\\
\hline
$\delta |V_{ud}|^2/|V_{ud}|^2$
&2&0&-2&0&0&$-0.0029\pm 0.0014$
\\
$\delta Q_W^{Cs}/Q_W^{Cs}$
&-4.82&5.41&0.05&0&0&$-0.0040\pm 0.0066$
\\
$\delta R_{e/\mu}$
&2&0&0&-2&0&$-0.0042 \pm 0.0033$
\\
$\delta G_\mu/G_\mu$
&0&0&1&0&0&$0.00025\pm 0.001875$
\\
$\delta R_{\nu}$
&0&0
&-0.21& 0.22& 0.08&$-0.0033\pm 0.0007$
\\
$\delta R_{\bar\nu}$
&0&0&-0.077& 0.132 &0.32
&$-0.0019\pm 0.0016$
\\
\hline
\end{tabular}
\caption{RPV contributions to $\delta |V_{ud}|^2/|V_{ud}|^2$, 
$\delta Q_W^{Cs}/Q_W^{Cs}$,
$\delta R_{e/\mu}$, $\delta G_\mu/G_\mu$, 
$\delta R_{\nu}$ and $\delta R_{\bar\nu}$.
Columns give the coefficients of the various corrections
$\Delta_{ijk}^{\prime}$
and $\Delta_{12k}$ to the different quantities.
The last column gives the experimentally  measured value of the corresponding
quantity.}
\label{table:RPV}
\end{table}

\begin{figure*}
\resizebox{8 cm}{!}{\includegraphics{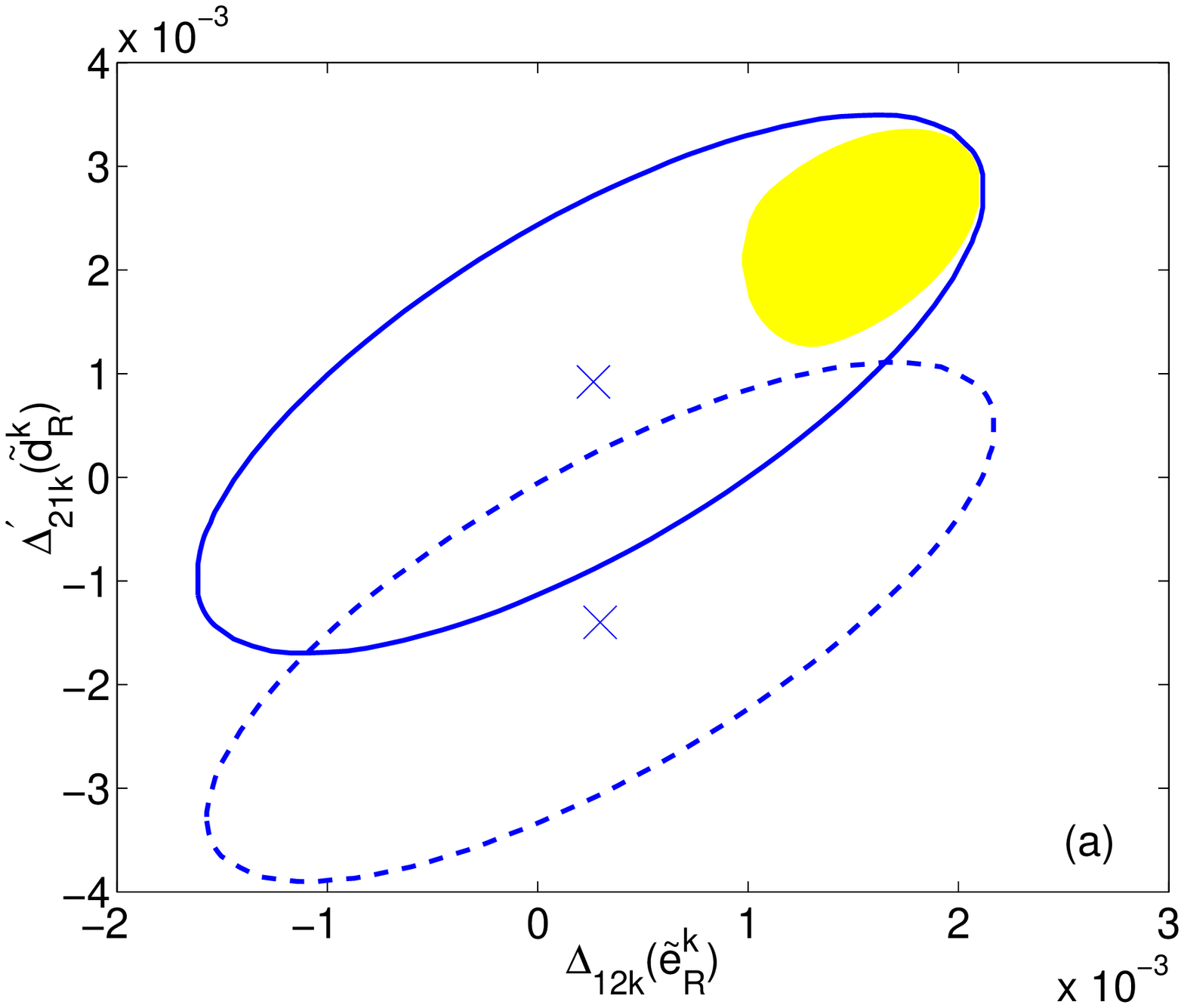}}
\resizebox{8 cm}{!}{\includegraphics{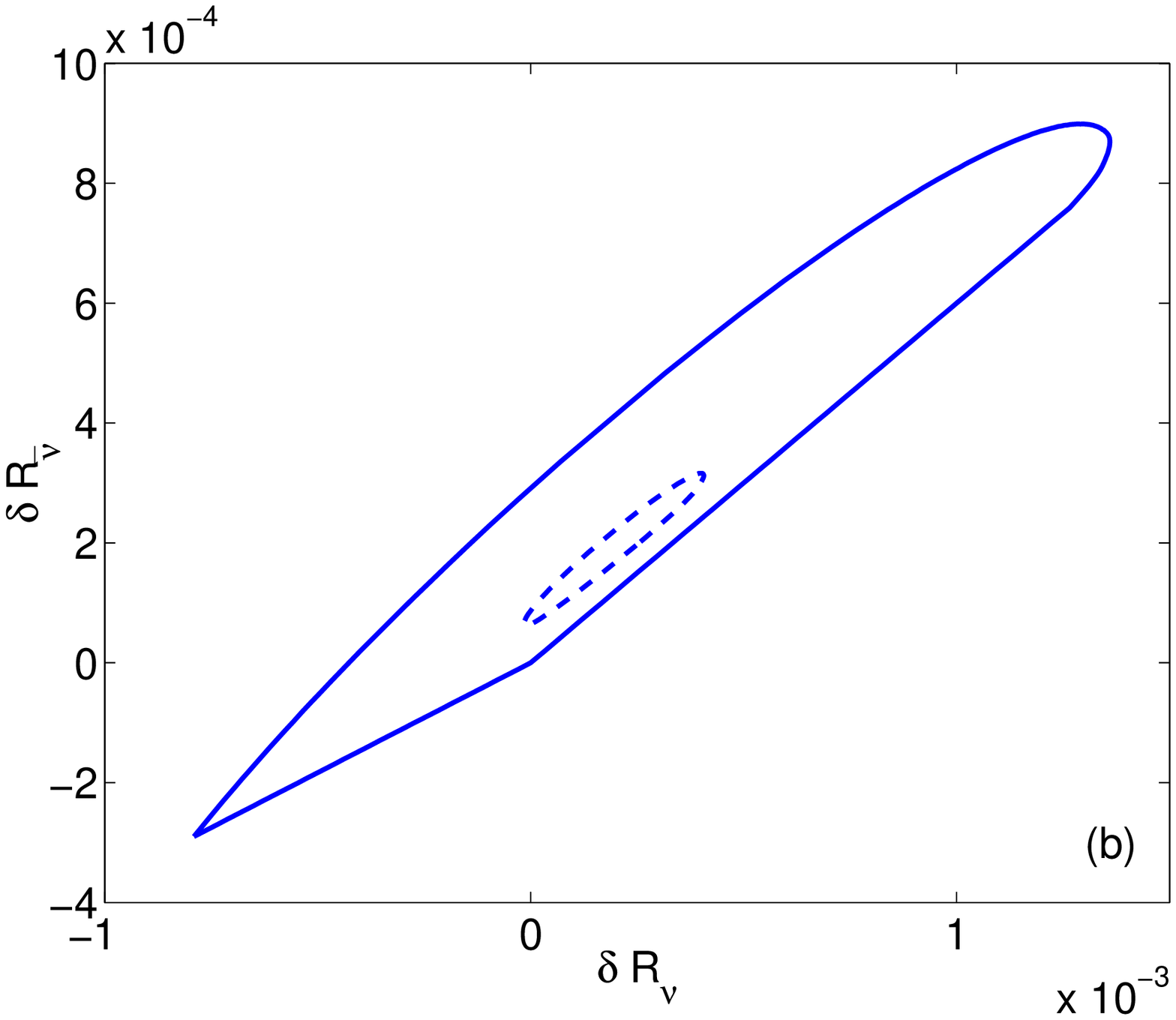}}
\caption{
Plot(a) shows the 1 $\sigma$ allowed region in
$\Delta_{12k}(\tilde{e}_R^k)$-$\Delta_{21k}^{\prime}(\tilde{d}_R^k)$
plane, with the best fit value denoted by the cross.  The solid (dashed)
ellipse is the fit excluding (including) the NuTeV 
$R_{\nu,\bar\nu}$ results.
Shading indicates the physically allowed region,
corresponding to $\Delta_{ijk}>0$ and $\Delta_{ijk}^{\prime}>0$.
Plot(b) shows the prediction for $\delta R_{\nu}$ and $\delta R_{\bar\nu}$,
using the 95$\%$ C.L. (solid line) or 1 $\sigma$ (dashed line)
allowed values for
$\Delta_{12k}(\tilde{e}_R^k)$ and $\Delta_{21k}^{\prime}(\tilde{d}_R^k)$
from fitting to $\delta |V_{ud}|^2/|V_{ud}|^2$, $\delta Q_W^{Cs}/Q_W^{Cs}$,
$\delta R_{e/\mu}$ and $\delta G_{\mu}/G_{\mu}$ with
$\Delta_{2k1}^{\prime}(\tilde{d}_L^k)$ set to zero.
}
\label{fig:fit}
\end{figure*}

Given the experimental constraints on the first four quantities in Table
\ref{table:RPV}, we obtain the one $\sigma$ allowed region for
$\Delta_{12k}(\tilde{e}_R^k)$ and $\Delta_{21k}^{\prime}(\tilde{d}_R^k)$
shown in Fig.~\ref{fig:fit} (a) by the solid 
ellipse.\footnote{In performing the
fit, we allow the signs of  $\Delta_{ijk}$, $\Delta_{ijk}^\prime$ to be
unconstrained.}
Since the RPV corrections $\Delta_{ijk}^{(\prime)}\propto
|\lambda_{ijk}^{(\prime)}|^2 \geq 0$,
the physically allowed region -- indicated by the shaded region in
Fig.~\ref{fig:fit} (a) --
corresponds to all of the $\Delta_{ijk}^{(\prime)}$ in
Table~\ref{table:RPV} being non-negative.
Taking the values of $\Delta_{12k}(\tilde{e}_R^k)$ and
$\Delta_{21k}^{\prime}(\tilde{d}_R^k)$
from this region, we obtain the allowed shifts in $\rnu$ and $\rnubar$
shown in Fig.~\ref{fig:fit} (b), dashed ellipse.
We also show the corresponding 95$\%$ C.L. region (solid line in 
Fig.~\ref{fig:fit} (b)).
For simplicity, we have set $\Delta_{2k1}^{\prime}(\tilde{d}_L^k)=0$ since
a non-zero value would yield only a positive
contribution to these quantities. Even so, the possible effects on $\rnu$
and $\rnubar$ are
by and large positive. While small negative corrections are also possible,
they are numerically
too small to be interesting.


We also performed a fit to the five RPV parameters including in addition
the NuTeV results for $\rnu$ and $\rnubar$.
The one $\sigma$ allowed region
for $\Delta_{12k}(\tilde{e}_R^k)$ and  $\Delta_{21k}^{\prime}(\tilde{d}_R^k)$
is given by the dashed
ellipse in Fig.~\ref{fig:fit} (a). At 95\% C.L., at least one of the
$\Delta_{ijk}^{(\prime)}$ must be negative in this fit. Thus, inclusion of the
NuTeV results appears to exclude the RPV SUSY effects summarized in Table
\ref{table:RPV}
with fairly high confidence.

\section{conclusion}

The NuTeV anomaly remains in need of explanation. If one is ultimately
unable to account
for it with conventional, SM effects ({\em e.g.}, small effects in parton
distribution functions), then one would require an explanation lying
outside the SM.
Here, we have shown that such an explanation would be hard to come by in
the MSSM alone.
In general, SUSY loop corrections to $\rnu$ and $\rnubar$ generally have
too small a magnitude
and the wrong sign to account for the effect. An exception occurs for
significant left-right
mixing among first generation squarks, where a sizable effect of the
necessary sign is
generated by gluino loops. At present, precision data exclude this scenario
with a high
degree of confidence, though a new analysis of $K_{\ell 3}$ decays may
weaken these
restrictions considerably. Even in this case, however, the value for the
weak mixing angle
extracted from neutrino-nucleus scattering will be largely unaffected by
gluino loops when
a Paschos-Wolfenstein type relation is used for the extraction.

R parity-violating effects also appear at odds with the NuTeV
anomaly. Inclusion of these effects could resolve an apparent conflict between
tests of charged current universality and implications of SUSY-breaking
models for the
MSSM spectrum\cite{kurylov02}. However they would also render the LSP unstable
and, therefore,
rule out SUSY dark matter. At face value, the NuTeV results appear to
disfavor this
resolution of the charged current universality problem. In short, we
conclude that the MSSM -- with or
without R-parity conservation -- is likely not responsible for the NuTeV
anomaly. The  culprit,
apparently, is to be found elsewhere.

\begin{acknowledgments}
We thank H. Schellman and J. Erler for useful discussions.
We thank Petr Vogel 
and Mark Wise for careful reading of the manuscript. 
This work is supported in part under U.S. Department of
Energy contract \#DE-FG03-02ER41215 (A.K. and M.J.R.-M.),  
\#DE-FG03-00ER41132 (M.J.R.-M.), and \#DE-FG03-92-ER-40701 (S.S.).
A.K. and M.J.R.-M. are supported by 
the National Science Foundation under award PHY00-71856.
S.S. is supported by the John A. McCone Fellowship.
\end{acknowledgments}

\appendix
\section{Essential Feynman rules}
\label{app:rules}

The complete set of Feynman rules for the MSSM 
can be found in \cite{haber,rosiek}. Here, we give only a brief
list of relevant vertices.  

The fermion-sfermion-gaugino vertices are:
\begin{center}
\mbox{
\resizebox{5 in}{!}{\includegraphics*[0,480][590,650]{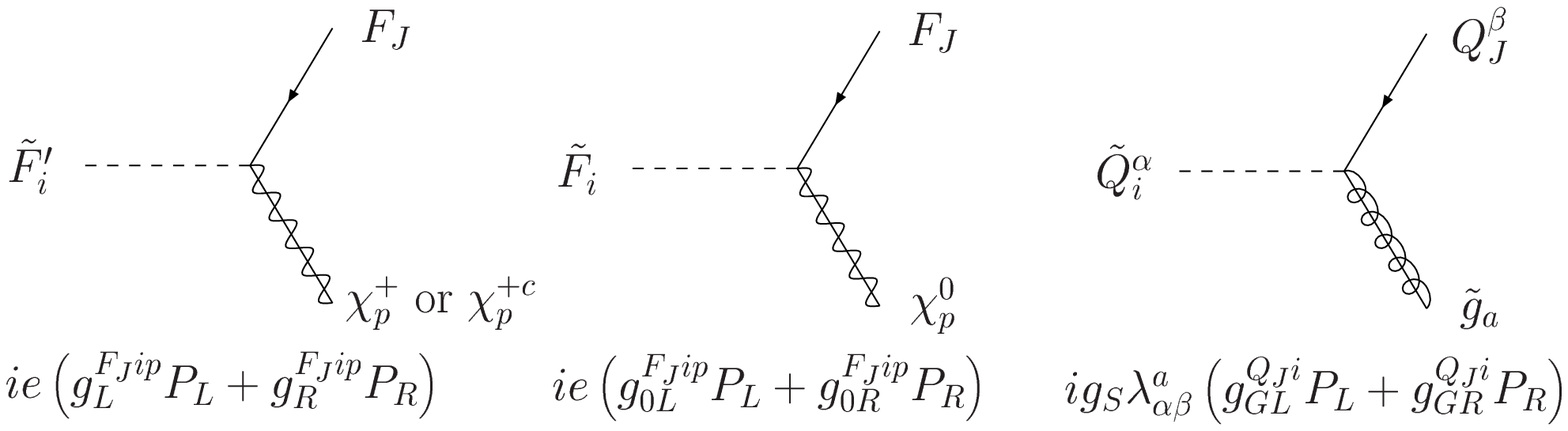}}
}
\end{center}
where $e$ is the absolute value
of the electron charge,  $g_S$ is the ${\rm SU(3)}_c$ coupling constant,
and $\lambda_{\alpha\beta}^a$ are Gell-Mann matrices \cite{rosiek}, 
normalized according to
$tr[\lambda^a \lambda^b]=1/2 \delta^{ab}$.
We use the capitalized letters $I$ and $J$ to denote 
the family index for quarks and leptons ($I,J=1,\cdots,3$), 
small letters $i$ and $j$ to denote the index for squarks and sleptons 
($i,j=1,\cdots,6$ except for sneutrino, when $i,j=1,\cdots,3$), and small 
letters $p$ and $n$ to denote the index for the neutralinos ($p,n=1,\cdots,4$)
and charginos ($p,n=1,2$). 
The index $a$ is reserved for the gluino index, $a=1,\cdots,8$.

The first vertex represents the coupling
of the fermion $F_J$ to the sfermion
$\tilde {F}^{\prime}_i$ and chargino $\chi_p^+$ (or its 
charge conjugation field $\chi_p^{+c}$ if $F=D,L$.). 
The chargino coupling constants $g^{F_Jip}_{L,R}$ 
for the quark sector are as follows (the repeated index is summed over):
\beqa
g_L^{U_Jip}&=&-\frac{1}{s}V_{IJ}^*\left(Z^{Ii}_DU_{p1}^*
-{m_{D_I} \over {\sqrt 2 M_W \cos\beta}}Z^{I+3,i}_DU_{p2}^* \right) 
\nonumber \\
g_R^{U_Jip}&=&\frac{1}{s}V_{IJ}^*{m_{U_I} \over 
{\sqrt 2 M_W \sin\beta}}Z^{Ii}_DU_{p2}
\nonumber \\
g_L^{D_Jip}&=&-\frac{1}{s}V_{IJ}\left(Z^{*Ii}_UV_{p1}^*
-{m_{U_I} \over {\sqrt 2 M_W \sin\beta}}Z^{*I+3,i}_UV_{p2}^* \right) 
\nonumber \\
g_R^{D_Jip}&=&\frac{1}{s}V_{IJ}{m_{D_I} \over 
{\sqrt 2 M_W \cos\beta}}Z^{*Ii}_UV_{p2},
\eeqa
where $s$ ($c$) is the sine (cosine) of Weinberg angle,
$V_{IJ}$ is the usual CKM matrix,
$m_{U_I},m_{D_I}$ is the mass of \lq\lq up" or \lq\lq down" quark
for generation I, $\tan\beta=v_u/v_d$ is the ratio of the
expectation values of the Higgs scalars, $Z_{U,D}$ are
the unitary 6$\times$6
mixing matrices  for up and down squarks, respectively,
that diagonalize full sfermion mass matrices, and
$U$ and $V$ are the unitary mixing matrices that diagonalize the 
chargino mass matrix
\cite{haber}. In practical calculations masses of the first and second 
generation quarks can be neglected as they are
much smaller than the mass of the $W$ boson. 
Similarly, for lepton sector,
\beqa
g_L^{\nu_Jip}&=&-\frac{1}{s}\left(Z^{Ji}_LU_{p1}^*
-{m_{L_J} \over {\sqrt 2 M_W \cos\beta}}Z^{J+3,i}_LU_{p2}^* \right) \nonumber \\
g_R^{\nu_Jip}&=&0 \nonumber \\
g_L^{L_Jip}&=&-\frac{1}{s}Z^{*Ji}_{\nu}V_{p1}^* \nonumber \\
g_R^{L_Jip}&=&\frac{1}{s}{m_{L_J} \over {\sqrt 2 M_W \cos\beta}}Z^{*Ji}_{\nu}V_{p2},
\eeqa
Note that the mixing matrix for sneutrinos $Z_{\nu}$ is 3$\times$3 because in the MSSM
neutrinos are purely left-handed.

The second vertex represents
coupling of a fermion to a sfermion and a neutralino.
The corresponding coupling constants $g_{0L,0R}^{F_Jip}$ for
the fermion-sfermion-neutralino vertex are:
\beqa
g_{0L}^{U_Jip}&=&-\frac{1}{\sqrt{2}sc}\left[Z^{*Ji}_U\left( N_{p2}^*c+{1\over 3}N_{p1}^*s\right)
+{m_{U_J} \over {M_Z \sin\beta}}Z^{*J+3,i}_UN_{p4}^* \right]\nonumber \\
g_{0R}^{U_Jip}&=&\frac{1}{\sqrt{2}c}\left({4\over 3}Z^{*J+3,i}_UN_{p1}
-{{m_{U_J}} \over {M_Z \sin\beta}}Z^{*Ji}_UN_{p4}\right) \nonumber \\
g_{0L}^{D_Jip}&=&\frac{1}{\sqrt{2}sc}\left[Z^{Ji}_D\left( N_{p2}^*c-{1\over 3}N_{p1}^*s\right)
-{m_{D_J} \over {M_Z \cos\beta}}Z^{J+3,i}_DN_{p3}^*\right] \nonumber \\
g_{0R}^{D_Jip}&=&-\frac{1}{\sqrt{2}c}\left[{2\over 3}Z^{J+3,i}_DN_{p1}
+{{m_{D_J}} \over {M_Z \cos\beta}}Z^{Ji}_DN_{p3}\right)\nonumber \\
g_{0L}^{\nu_Jip}&=&-\frac{1}{\sqrt{2}sc}Z^{*Ji}_{\nu}\left( N_{p2}^*c-N_{p1}^*s\right] \nonumber \\
g_{0R}^{\nu_Jip}&=&0 \nonumber \\
g_{0L}^{L_Jip}&=&\frac{1}{\sqrt{2}sc}\left[Z^{Ji}_L\left( N_{p2}^*c+N_{p1}^*s\right)
-{m_{L_J} \over {M_Z \cos\beta}}Z^{J+3,i}_LN_{p3}^*\right] \nonumber \\
g_{0R}^{L_Jip}&=&-\frac{1}{\sqrt{2}c}\left(2Z^{J+3,i}_LN_{p1}
+{{m_{L_J}} \over {M_Z \cos\beta}}Z^{Ji}_LN_{p3}\right) ,
\eeqa
where $N$ is a 4$\times$4 mixing matrix that diagonalizes 
neutralino mass matrix\cite{haber}. 

Finally, quark-squark-gluino (the third vertex)
couplings $g_{GL,GR}^{F_Ji}$ are:
\beqa
g_{GL}^{U_Ji}=-\sqrt{2}Z^{*Ji}_U &\ \ \ & 
g_{GR}^{U_Ji}=\sqrt{2}Z^{*J+3,i}_U \nonumber \\
g_{GL}^{D_Ji}=-\sqrt{2}Z^{Ji}_D &\ \ \ &
g_{GR}^{D_Ji}=\sqrt{2}Z^{J+3,i}_D.
\eeqa

The gauge boson-gaugino-gaugino couplings are given as follows:
\begin{center}
\mbox{
\resizebox{5. in}{!}{\includegraphics*[0,480][580,660]{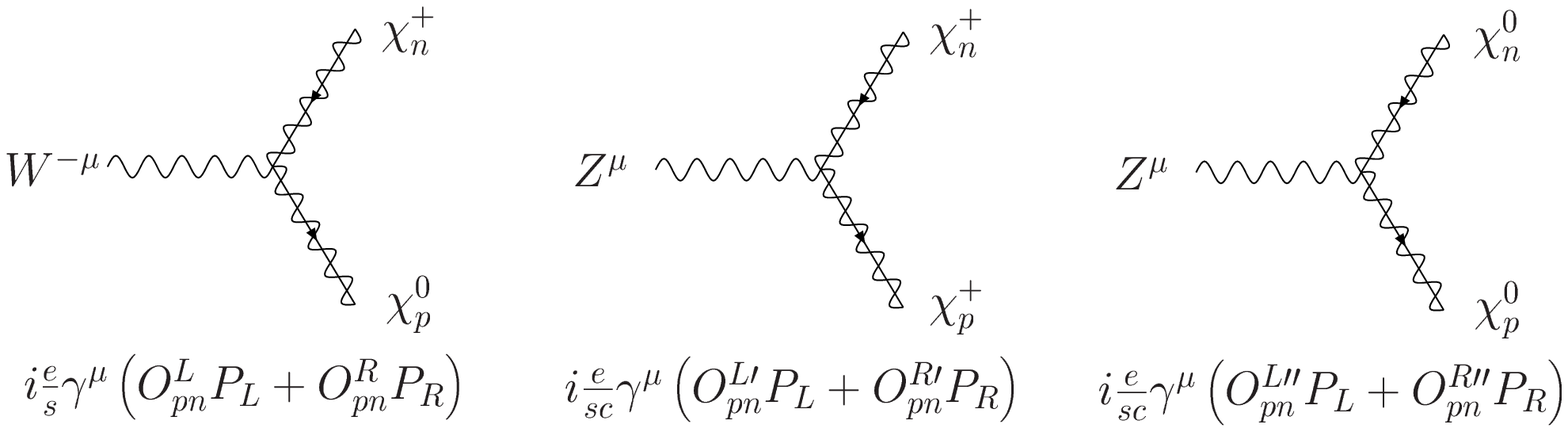}}
}
\end{center}
where 
\beqa
\label{eq:O-matrices-W}
O^L_{pn}=\left( N_{p2}V^*_{n1}-{1\over{\sqrt 2}} N_{p4}V^*_{n2} \right)
&\ \ \ &
O^R_{pn}=\left( N^*_{p2}U_{n1}+{1\over{\sqrt 2}} N^*_{p3}U_{n2} \right)
\nonumber \\
O^{L\prime}_{pn}=
\left(-V_{p1}V^*_{n1}-{1\over 2} V_{p2}V^*_{n2}+\delta_{pn}s^2\right) 
&\ \ \ &
O^{R\prime}_{pn}=
\left(-U^*_{p1}U_{n1}-{1\over 2} U^*_{p2}U_{n2}+\delta_{pn}s^2\right) \nonumber \\
O^{L\prime\prime}_{pn}=
\left(-{1\over 2}N_{p3}N^*_{n3}+{1\over 2}N_{p4}N^*_{n4} \right) 
&\ \ \ &
O^{R\prime\prime}_{pn}=-O^{*L\prime\prime}_{pn}.
\eeqa
In addition, for the photon-chargino-chargino vertex  
$A^{\mu}-\chi_n^+-\chi_p^+$ is given by 
$-ie\gamma^\mu\delta^{pn}$.

The gauge boson-sfermion-sfermion vertices are given by:
\begin{center}
\mbox{
\resizebox{4. in}{!}{\includegraphics*[0,480][420,670]{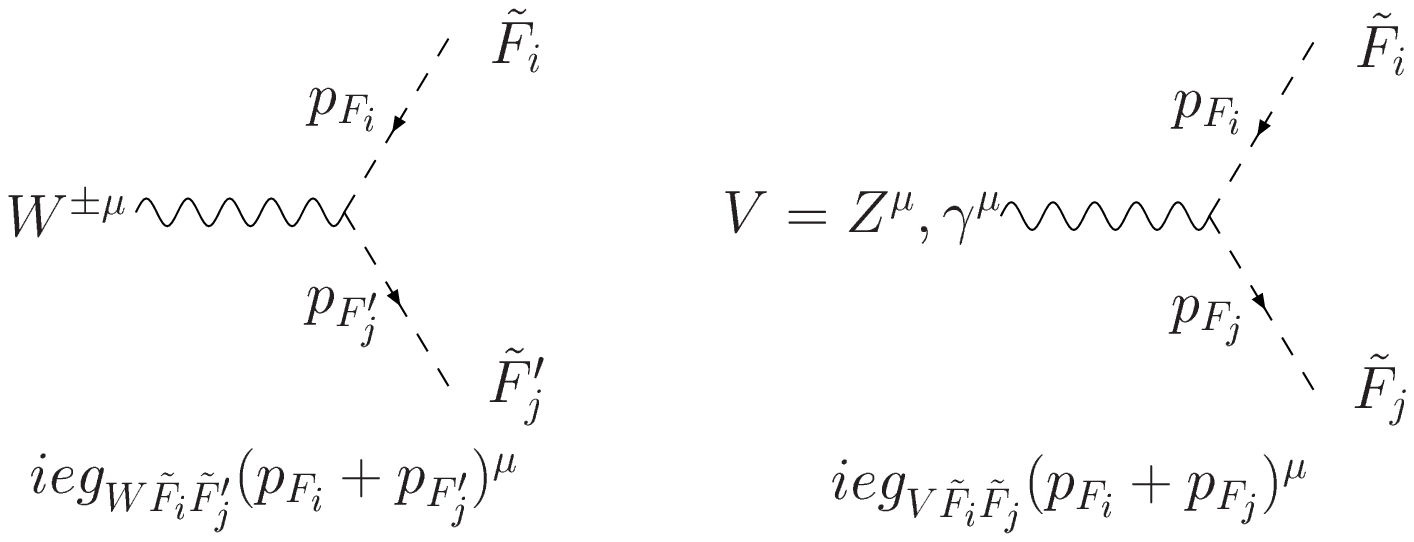}}
}
\end{center}
For the $W$- coupling:
\beq
g_{W\tilde{D}_i\tilde{U}_j}=-\frac{1}{\sqrt{2}s}V_{JI}Z_D^{*Ii}Z_U^{*Jj},
\ \ \ 
g_{W\tilde{L}_i\tilde{\nu}_j}=-\frac{1}{\sqrt{2}s}Z_L^{*Ii}Z_{\nu}^{*Jj} .
\eeq
and for the $Z$ and $\gamma$ couplings:
\beqa
g_{Z\tilde{F}_i\tilde{F}_j}&=&-\frac{1}{sc}\left(I_3^F Z_F^{Ii}Z_F^{*Ij}
-Q_Fs^2\delta_{ij}\right) \nonumber \\
g_{\gamma\tilde{F}_i\tilde{F}_j}&=&-Q_F \delta^{ij}.
\eeqa
Here, $I_3^F$ and $Q_F$ are the isospin and the 
electric charge of the sfermions $\tilde{F}$, respectively.

The two scalar-two gauge boson vertices necessary 
for the calculation of the gauge
boson self energies are (see Ref.~\cite{rosiek}):
\begin{center}
\mbox{
\resizebox{3. in}{!}{\includegraphics*[10,570][320,650]{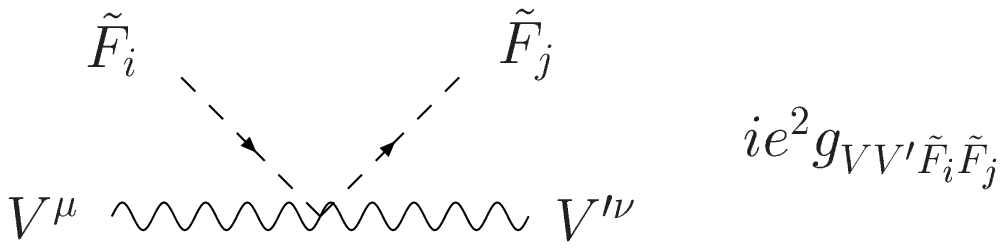}}
}
\end{center}
where 
\beqa
g_{WW\tilde{F}_i\tilde{F}_j}&=&\frac{1}{2s^2}Z_F^{Ki}Z_F^{*Kj}g^{\mu\nu}
\nonumber \\
g_{ZZ\tilde{F}_i\tilde{F}_j}&=&
\frac{1}{s^2c^2}
\left(2I_3^F(I_3^F-2Q_F s^2)Z_F^{Ki}Z_F^{*Kj}
+2 Q_F^2 s^4 \delta_{ij} \right)g^{\mu\nu} 
\nonumber \\
g_{Z \gamma \tilde{F}_i\tilde{F}_j}&=&
\frac{1}{sc}
Q_F\left(2I_3^F Z_F^{Ki}Z_F^{*Kj}-2Q_F s^2 \delta_{ij}\right)g^{\mu\nu}
\nonumber \\
g_{\gamma\gamma\tilde{F}_i\tilde{F}_j}&=&
2Q_F^2\delta_{ij}g^{\mu\nu} .
\eeqa
%


\section{Expressions for the relevant Feynman diagrams}
\label{app:expressions}
In this section, we list analytical expressions for the MSSM 
contribution to
the gauge boson self-energies, the fermion wave function renormalization,
the gauge boson-fermion-fermion vertex correction, and the box 
diagrams relevant to the neutrino-nucleus scatterings. Modified dimensional
reduction (${\overline{DR}}$) scheme is used in computing MSSM loop
contributions, although in the Appendices, we have neglected the hat 
in all relevant variables.

\subsection{Gauge Boson Self-Energies}

We first define the following class of two-point integration functions:
\beqa
\label{eq:F-functions}
F_n(m_1,m_2,m_3)&=&{\int_0^1}x^n\ln
\left\{\left[x m_1^2+(1-x)m_2^2-x(1-x)m_3^2\right]/\mu^2 \right\},
\\
F_{12}(m_1,m_2,m_3)&=&F_1(m_1,m_2,m_3)-F_2(m_1,m_2,m_3) ,
\eeqa
where $\mu$ is the renormalization scale. 
In our analysis, $m_3^2$ will be replaced with either the external 
fermion mass squared
or one of the Mandelstaam variables. For the
first two generation quarks and the first two generation leptons the fermion
mass can be set to zero. Focusing on the case when all Mandelstaam variables
are small compared to $m_{1,2}^2$, we obtain:
\beqa
F_0(m_1,m_2,0)&=&\ln m_1^2-1+{m_2^2\over{m_2^2-m_1^2}}\ln{m_2^2\over m_1^2}
-\ln \mu^2
 \nonumber \\
F_1(m_1,m_2,0)&=&{1\over 4}\Biggl[2\ln m_1^2-1+{2m_2^2\over{m_1^2-m_2^2}}
+{2m_2^4\over{(m_2^2-m_1^2)^2}}\ln{m_2^2\over m_1^2
}-2 \ln \mu^2\Biggr].
\eeqa

\subsubsection{Gaugino Loops}

\begin{figure}
\resizebox{16. cm}{!}{\includegraphics*[10,470][580,610]{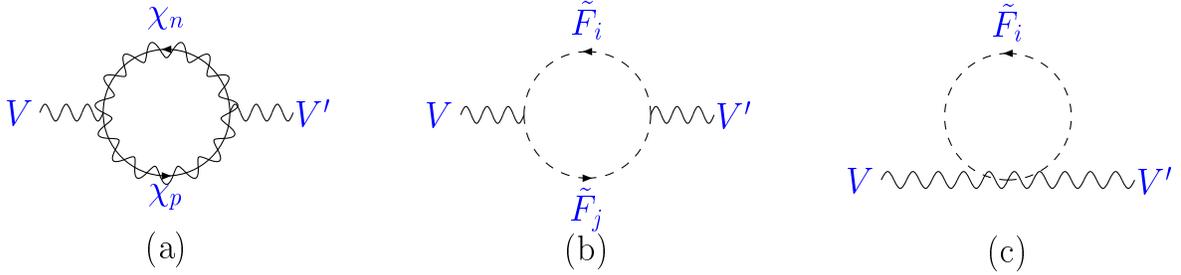}}
\caption{Feynman diagrams of one loop SUSY contributions to the
gauge boson self energy.}
\label{fig:wave_boson}
\end{figure}

The Feynman diagrams of this type are shown in Fig. \ref{fig:wave_boson}(a).
The contribution of gaugino loops to $W$ and $Z$ self-energies is given by
\beqa
\Pi_{VV}^{\chi\chi}(q^2)&=&-{\alpha\over 2\pi}\sum_{p,n}
\Biggl\{
(g_{L}g_{R}^*+g_L^*g_R)m_{\chi_n}m_{\chi_p}F_0(m_{\chi_n},m_{\chi_p},q)
\nonumber \\
&&\hspace*{-1 in}+\left(|g_L|^2+|g_R|^2\right) \left[2q^2 
F_{12}(m_{\chi_n},m_{\chi_p},q)
-m_{\chi_n}^2 F_1(m_{\chi_n},m_{\chi_p},q)
-m_{\chi_p}^2 F_1(m_{\chi_p},m_{\chi_n},q)
\right]
\Biggr\},
\eeqa
where $q^\mu$ is the momentum carried by gauge boson $V$, and 
$q^2=q^\mu q_\mu$.  The couplings $g_L$ and $g_R$ are 
listed below for $\Pi_{WW}$
and $\Pi_{ZZ}$:
\begin{center}
\begin{tabular}{|c|c|c|c|c|c|}\hline
$\Pi_{VV}$&$\chi_p$&$\chi_n$&$g_L$&$g_R$&comment\\\hline
$\Pi_{WW}^{\chi^0\chi^+}$ & $\chi^0_p$&$\chi^+_n$&$O^L_{pn}/s$ &$O^R_{pn}/s$&\\
$\Pi_{ZZ}^{\chi^+\chi^+}$ & $\chi^+_p$&$\chi^+_n$&$O^{L\prime}_{pn}/sc$ 
&$O^{R\prime}_{pn}/sc$&\\
$\Pi_{ZZ}^{\chi^0\chi^0}$ & $\chi^0_p$&$\chi^0_n$&$O^{L\prime\prime}_{pn}/sc$ 
&$O^{R\prime\prime}_{pn}/sc$&multiply $1/2$\\
\hline
\end{tabular}
\end{center}

The $Z-\gamma$ mixing tensor 
and photon self-energy from chargino loops is given by
\beqa
\Pi_{V V^{\prime}}^{\chi^+}(q^2)&=&-q^2{\alpha\over \pi} \sum_p
g_V g_{V\prime} F_{12}(m_{\chi^+_p},m_{\chi^+_p},q), 
\eeqa
with the couplings $g_V$ and $g_{V^{\prime}}$ listed below:	
\begin{center}
\begin{tabular}{|c|c|c|c|}\hline
$\Pi_{VV^{\prime}}$&$g_V$&$g_{V\prime}$&comment\\\hline
$\Pi_{Z \gamma }^{\chi^+}$ &$(O^{L\prime}_{pp}+O^{R\prime}_{pp})/sc$&-1&\\
$\Pi_{\gamma \gamma}^{\chi^+}$ &-1&-1&
multiply $2$\\
\hline
\end{tabular}
\end{center}

\subsubsection{Scalar Loops}

Diagrams of this type are shown in Figs. \ref{fig:wave_boson}(b,c). 
The contribution of  Fig. \ref{fig:wave_boson} (b) to 
$\Pi_{WW}$, 
$\Pi_{ZZ}$, $\Pi_{Z \gamma}$ and  $\Pi_{\gamma\gamma}$ is given by 
\beqa
\label{eq:ww-slepton-sneut}
\Pi_{VV^{\prime}}^{{\tilde F}{\tilde F} }(q^2)
&=&-{\alpha \over 4 \pi}\sum_{i,j}
g_{V\tilde{F}_i\tilde{F}_j} 
g_{V^{\prime} \tilde{F}_i\tilde{F}_j}^*
\Biggl\{
m_{{\tilde F}_i}^2+m_{{\tilde F}_j}^2-{q^2\over 3} \nonumber \\
&&-2\Biggl[
m_{{\tilde F}_i}^2 F_1(m_{{\tilde F}_i},m_{{\tilde F}_j},q)
+m_{{\tilde F}_j}^2 F_1(m_{{\tilde F}_j},m_{{\tilde F}_i},q) -q^2F_{12}(m_{{\tilde F}_i},m_{{\tilde F}_j},q)
\Biggr]
\Biggr\}.
\label{eq:PIVV}
\eeqa
The sfermion pair $(\tilde{F}_i, \tilde{F}_j)$ running in the loop for 
$\Pi_{V V^{\prime}}$ is listed below:
\begin{center}
\begin{tabular}{|c|c|}\hline
$(V, V^{\prime})$&$(\tilde{F}_i, \tilde{F}_j)$\\\hline
$(W,W)$& $(\tilde{L}_i, \tilde{\nu}_j)$, $(\tilde{D}_i, \tilde{U}_j)$
\\
$(Z,Z)$& $(\tilde{\nu}_i, \tilde{\nu}_j)$,
$(\tilde{L}_i, \tilde{L}_j)$, $(\tilde{U}_i, \tilde{U}_j)$
,$(\tilde{D}_i, \tilde{D}_j)$ 
\\
$(Z,\gamma)$&$(\tilde{L}_i, \tilde{L}_{j})$,
$(\tilde{U}_i, \tilde{U}_{j})$ ,$(\tilde{D}_i, \tilde{D}_{j})$ 
\\
$(\gamma, \gamma)$&$(\tilde{L}_i, \tilde{L}_{j})$,
$(\tilde{U}_i, \tilde{U}_{j})$ ,$(\tilde{D}_i, \tilde{D}_{j})$ 
\\
\hline
\end{tabular}
\end{center}
For squark contributions, an 
additional color factor of $N_c$ enters the right-hand side of 
Eq.~(\ref{eq:PIVV}).

The contribution of Fig. \ref{fig:wave_boson} (c) 
to the self-energies is given by 
\beqa
\Pi_{VV^{\prime}}^{{\tilde F}}(q^2)&=&{\alpha \over 4 \pi }\sum_{i}
g_{V V^{\prime} \tilde{F}_i \tilde{F}_i}
m_{{\tilde F}_i}^2 \left(1-\ln m_{{\tilde F}_i}^2 + \ln \mu^2\right),
\eeqa
where $\tilde{F}=\tilde\nu,\ \tilde{L}, \tilde{U}$ and $\tilde{D}$ except
for $V^{\prime}=\gamma$, for which $\tilde{\nu}$ does not contribute. 

When $q^2\rightarrow 0$, $\Pi_{Z \gamma }(q^2)/q^2$ and 
$\Pi_{\gamma \gamma}(q^2)/q^2$ from sfermion loops reduce to
\beq
\frac{\Pi_{VV^{\prime}}(q^2)}{q^2}=-\frac{\alpha}{12\pi}\sum_{i}
g_{V\tilde{F}_i\tilde{F}_i} 
g_{V^{\prime} \tilde{F}_i\tilde{F}_i}^*
\left(\ln m^2_{\tilde{F}_i}-\ln \mu^2 \right).
\eeq

\subsection{Field Strength Renormalization for Fermions}

\begin{figure}
\resizebox{5. cm}{!}{\includegraphics*[30,520][160,610]{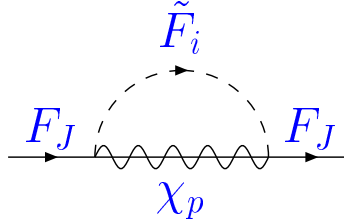}}
\caption{Feynman diagrams of one loop SUSY contribution to the
fermion wave function
renormalization.}
\label{fig:wave_fer}
\end{figure}

The diagrams contributing to the field strength renormalization  for
fermions all have the form shown in Fig \ref{fig:wave_fer}. Therefore,
the same formula can be applied to contributions with charginos and
neutralinos in the loop, provided that the appropriate couplings and
masses are used. The gluino contributions are multiplied by an
extra factor of Casimir factor $C_2(N)=4/3$ for ${\rm SU(3)}_c$. The
field strength renormalization for the left-handed quark $F_J$ is
\beqa
\label{eq:Z-quarks}
\delta Z^{F_J}_L&=&{{\alpha}\over{4\pi}} {\sum_{i,p}}
{\left|g^{F_Jip}_L\right|}^2 
F_1(m_{{\tilde F^{\prime}}_i},m_{\chi^+_p},m_{F_J}) 
+{{\alpha}\over{4\pi}} {\sum_{i,p}}
{\left|g^{F_Jip}_{0L}\right|}^2 F_1(m_{{\tilde F}_i},m_{\chi^0_p},m_{F_J}) 
\nonumber \\
&&+{4 \over 3} \times {{\alpha_S}\over{4\pi}} {\sum_{i}}
{\left|g^{F_Ji}_{GL}\right|}^2 F_1(m_{{\tilde F}_i},m_{\tilde{g}},m_{F_J}) , 
\eeqa
where $m_{\tilde{g}}$ is the gluino mass and 
$F^{\prime}$ stands for the isopartner of the 
fermion $F$ ({\it e.g.} if $F$ is the up-type quark, then $F^{\prime}$ 
is the down-type quark).
By replacing $L\to R$ one easily obtains the field strength 
renormalization for the right-handed
fermions. The same formulae apply to the sleptons provided 
that the appropriate couplings and masses
are used. Naturally, corrections involving the strong coupling are absent in that case.


\subsection{Vertex Corrections}

\begin{figure}
\resizebox{10. cm}{!}{\includegraphics*[30,510][370,640]{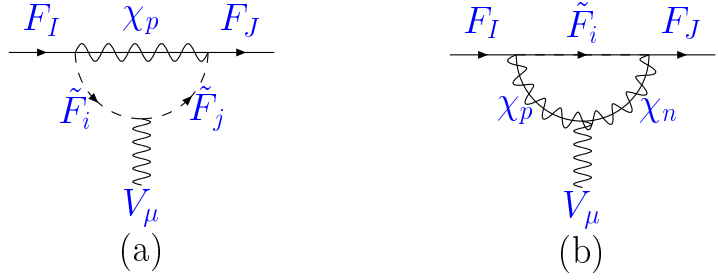}}
\caption{Feynman diagrams of one loop SUSY contribution to the
gauge coupling vertex.}
\label{fig:vertex}
\end{figure}

One-loop SUSY corrections to the $V-f-f^\prime$ vertex are shown in
Fig. \ref{fig:vertex}. There are two types of corrections: 
loops with the
vector boson coupling to the scalar particles,
Fig. \ref{fig:vertex}(a), and 
loops with the vector bosons coupling to the
gauginos, Fig. \ref{fig:vertex}(b).  The complete set of Feynman
diagrams for each vertex, such as $W-d-u$, can contain more than one
diagram of each type. However, since these diagrams differ only in the
specific values of the masses and couplings needed to obtain the
numerical answer we do not show all graphs explicitly.  Below, we use
the superscripts (a) and (b) to distinguish between the two types of
loop graphs. To distinguish the graphs of the same type, we supplement
the superscript with a number.

We first define the three-point 
integration functions needed for the evaluation of
the triangle diagrams:
\beqa
V_1(M,m_1,m_2)&=&\int_0^1dx\int_0^1dy{y\over D_3(M,m_1,m_2)}\nonumber \\
V_2(M,m_1,m_2)&=&\int_0^1dx\int_0^1dy{y\ln 
\left[D_3(M,m_1,m_2)/\mu^2\right]\nonumber} \\
D_3(M,m_1,m_2)&=&(1-y)M^2+y[(1-x)m_1^2+xm_2^2] .
\eeqa
Explicitly:
\beqa
\label{eq:vert-functions}
V_1(M,m_1,m_2)&=&{{m_1^2\ln{m_1^2\over M^2}}
\over{(M^2-m_1^2)(m_2^2-m_1^2)}} 
+{{m_2^2\ln{m_2^2\over M^2}}
\over{(M^2-m_2^2)(m_1^2-m_2^2)}}
\nonumber \\
V_2(M,m_1,m_2)&=&{1\over 4}\Biggl[2\ln{M^2}-3
+{2m_1^4\over{(M^2-m_1^2)(m_2^2-m_1^2)}}\ln{m_1^2\over M^2}
\nonumber \\
&&+{2m_2^4\over{(M^2-m_2^2)(m_1^2-m_2^2)}}\ln{m_2^2\over M^2} 
-2 \ln \mu^2\Biggr] .
\eeqa
Defining the matrix element for the vertex to be 
\beq
{\cal{M}}=ie [\delta_{V F_I F_J}^L {\bar F_J} \gamma^\mu P_L F_I
+ \delta_{V F_I F_J}^R {\bar F_J} \gamma^\mu P_R F_I] V_{\mu} .
\eeq
Diagram (a) gives
\beqa
\delta_{V F_IF_J}^{L;(a)}&=&-{{\alpha}\over{4 \pi }}{\sum_{p,i,j}}
g_{V \tilde{F}_i \tilde{F}_j}
g_{\chi L}^{I}g_{\chi L}^{J*}
V_2(m_{\chi_p},m_{{\tilde F}_i},m_{{\tilde F}_j})\\
\delta_{V F_IF_J}^{R;(a)}&=&-{{\alpha}\over{4 \pi }}{\sum_{p,i,j}}
g_{V \tilde{F}_i \tilde{F}_j}
g_{\chi R}^{I}g_{\chi R}^{J*}
V_2(m_{\chi_p},m_{{\tilde F}_i},m_{{\tilde F}_j}) .
\eeqa
Diagram (b) gives
\beqa
\delta_{V F_IF_J}^{L;(b)}&=&{{\alpha}\over{4 \pi}}{\sum_{p,n,i}}
\left[ g_{V\chi\chi}^L g_{\chi L}^{I}g_{\chi L}^{J*}
m_{\chi_p}m_{\chi_n}V_1 
-g_{V\chi\chi}^R g_{\chi L}^{I}g_{\chi L}^{J*} \left(1/2+V_2
\right) \right]
\label{eq:VFF1}
\\
\delta_{V F_IF_J}^{R;(b)}&=&{{\alpha}\over{4 \pi}}{\sum_{p,n,i}}\left[ 
g_{V\chi\chi}^R g_{\chi R}^{I}g_{\chi R}^{J*}
m_{\chi_p}m_{\chi_n}V_1
-g_{V\chi\chi}^L g_{\chi R}^{I}g_{\chi R}^{J*}\left(1/2+ V_2\right)
\right],
\label{eq:VFF2}
\eeqa
with the argument for functions $V$'s in Eqs.~(\ref{eq:VFF1}), (\ref{eq:VFF2})
being $V_{1,2}(m_{\tilde F_i},m_{\chi_p},m_{\chi_n})$.
The explicit expression for $g_{V\tilde{F}_i\tilde{F}_j}$, 
$g_{V\chi\chi}^{L,R}$, $g_{\chi L}^{I,J}$
and $g_{\chi R}^{I,J}$ for each individual vertex 
diagrams will be given below. 

\subsubsection{Charged Current Vertex}
The vertex correction to $W-D_I-U_J$ is 
\begin{center}
\begin{tabular}{|c|ccc|c|c|c||c|ccc|c|c|c|c|}\hline
$\delta_{V F_IF_J}^{(a)}$&$\chi_p$&$\tilde{F}_i$&$\tilde{F}_j$&
$g_{V\tilde{F}_i\tilde{F}_j}$&$g_{\chi L,R}^I$&$g_{\chi L,R}^J$
&$\delta_{V F_IF_J}^{(b)}$&$\chi_p$&$\chi_n$&$\tilde{F}_i$
&$g_{V\chi\chi}^L$&$g_{V\chi\chi}^R$
&$g_{\chi L,R}^I$&$g_{\chi L,R}^J$
\\\hline
$\delta_{W D_IU_J}^{(a1)}$&$\chi^0_p$&$\tilde{D}_i$&$\tilde{U}_j$&
$g_{W \tilde{D}_i\tilde{U}_j}$&$g_{0L,R}^{D_I ip}$&$g_{0L,R}^{U_J jp}$
&$\delta_{W D_IU_J}^{(b1)}$&$\chi^0_p$&$\chi^+_n$&$\tilde{D}_i$
&$O_{pn}^{*L}/s$&$O_{pn}^{*R}/s$&$g_{0L,R}^{D_Iip}$&$g_{L,R}^{U_Jin}$\\
&&&&&&&
$\delta_{W D_IU_J}^{(b2)}$&$\chi^+_p$&$\chi^0_n$&$\tilde{U}_i$
&$-O_{np}^{*R}/s$&$-O_{np}^{*L}/s$&$g_{L,R}^{D_Iip}$&$g_{0L,R}^{U_Jin}$\\

\hline
\end{tabular}
\end{center}
 
The vertex correction due to gluino exchange is similar to 
$\delta_{WD_IU_J}^{(a1)}$, 
with the substitution of 
\beqa
&&\alpha\rightarrow \alpha_S,\ \ \  
\chi^0_p \rightarrow \tilde{g},\ \ \  
g_{0L,R}^{D_I ip} \rightarrow g_{GL,R}^{D_I i},\ \ \ 
g_{0L,R}^{U_J jp} \rightarrow g_{GL,R}^{U_J j}, \ \ \ 
\label{eq:gluinosub}
\eeqa
and multiplication of  the whole expression by 4/3. 
This substitution rule also applies for the neutral current vertex 
listed in the next section. 

Similar to the case of the field strengths, 
the corresponding vertex corrections involving leptons $\delta_{V L_I\nu_J}$
are obtained from by using the appropriate masses and couplings.
The gluino loop corrections must be omitted in that case.

\subsubsection{Neutral Current Vertex}

For the $Z-D_I-D_J$ vertex
\begin{center}
\begin{tabular}{|c|ccc|c|c|c||c|ccc|c|c|c|c|}\hline
$\delta_{V F_IF_J}^{(a)}$&$\chi_p$&$\tilde{F}_i$&$\tilde{F}_j$&
$g_{V\tilde{F}_i\tilde{F}_j}$&$g_{\chi L,R}^I$&$g_{\chi L,R}^J$
&$\delta_{V F_IF_J}^{(b)}$
&$\chi_p$&$\chi_n$&$\tilde{F}_i$&$g_{V\chi\chi}^L$&$g_{V\chi\chi}^R$
&$g_{\chi L,R}^I$&$g_{\chi L,R}^J$
\\\hline
$\delta_{Z D_ID_J}^{(a1)}$&$\chi^0_p$&$\tilde{D}_i$&$\tilde{D}_j$&
$g_{Z \tilde{D}_i\tilde{D}_j}$&$g_{0L,R}^{D_I ip}$&$g_{0L,R}^{D_J jp}$
&$\delta_{ZD_ID_J}^{(b1)}$&$\chi^0_p$&$\chi^0_n$&$\tilde{D}_i$
&$O_{np}^{L\prime\prime}/sc$&$O_{np}^{R\prime\prime}/sc$&$g_{0L,R}^{D_Iip}$&$g_{0L,R}^{D_Jin}$\\
$\delta_{ZD_ID_J}^{(a2)}$&$\chi^+_p$&$\tilde{U}_i$&$\tilde{U}_j$&
$g_{Z \tilde{U}_i\tilde{U}_j}$&$g_{L,R}^{D_I ip}$&$g_{L,R}^{D_J jp}$
&
$\delta_{ZD_ID_J}^{(b2)}$&$\chi^+_p$&$\chi^+_n$&$\tilde{U}_i$
&$-O_{pn}^{R\prime}/sc$&$-O_{pn}^{L\prime}/sc$&$g_{L,R}^{D_Iip}$&$g_{L,R}^{D_Jin}$\\
\hline
\end{tabular}
\end{center}

For the $Z-U_I-U_J$ vertex
\begin{center}
\begin{tabular}{|c|ccc|c|c|c||c|ccc|c|c|c|c|}\hline
$\delta_{VF_IF_J}^{(a)}$&$\chi_p$&$\tilde{F}_i$&$\tilde{F}_j$&
$g_{V\tilde{F}_i\tilde{F}_j}$&$g_{\chi L,R}^I$&$g_{\chi L,R}^J$
&$\delta_{VF_IF_J}^{(b)}$
&$\chi_p$&$\chi_n$&$\tilde{F}_i$&$g_{V\chi\chi}^L$&$g_{V\chi\chi}^R$
&$g_{\chi L,R}^I$&$g_{\chi L,R}^J$
\\\hline
$\delta_{ZU_IU_J}^{(a1)}$&$\chi^0_p$&$\tilde{U}_i$&$\tilde{U}_j$&
$g_{Z \tilde{U}_i\tilde{U}_j}$&$g_{0L,R}^{U_I ip}$&$g_{0L,R}^{U_J jp}$
&$\delta_{Z U_IU_J}^{(b1)}$&$\chi^0_p$&$\chi^0_n$&$\tilde{U}_i$
&$O_{np}^{L\prime\prime}/sc$&$O_{np}^{R\prime\prime}/sc$&$g_{0L,R}^{U_Iip}$&$g_{0L,R}^{U_Jin}$\\
$\delta_{ZU_IU_J}^{(a2)}$&$\chi^+_p$&$\tilde{D}_i$&$\tilde{D}_j$&
$g_{Z \tilde{D}_i\tilde{D}_j}$&$g_{L,R}^{U_I ip}$&$g_{L,R}^{U_J jp}$
&
$\delta_{ZU_IU_J}^{(b2)}$&$\chi^+_p$&$\chi^+_n$&$\tilde{D}_i$
&$O_{np}^{L\prime}/sc$&$O_{np}^{R\prime}/sc$
&$g_{L,R}^{U_Iip}$&$g_{L,R}^{U_Jin}$\\
\hline
\end{tabular}
\end{center}

The radiative corrections to the lepton neutral current vertex are
directly obtained from the above expressions by replacing the up 
(s)quark with the
(s)neutrino and the down (s)quark with the (s)electron.

\subsection{Anapole Moment Corrections}

In the presence of parity-violating interactions, higher-order contributions 
can generate the photon-fermion-fermion coupling of the form (see, e.g. 
Ref.~\cite{michael}):
\beq
{\cal{L}}_{\gamma ff}^{PV}=-e \frac{F_{A,f}(q^2)}{M_Z^2} 
\bar f(q^2 \gamma^\mu -
\slashq q^\mu) \gamma_5 f A_{\mu} .
\eeq
The contributions 
from gaugino-sfermion loop at $q^2 \rightarrow 0$ are:
\beqa
{F}_{A,f}(q^2)&=&{F}_{A,f}^{(a1)}(q^2)+{F}_{A,f}^{(a2)}(q^2)
+{F}_{A,f}^{(b)}(q^2)
\nonumber \\
{F}_{A,f}^{(a1)}(q^2)&=&-Q_f {\alpha M_Z^2 \over{48\pi }}\sum_{i,p}
\left(|g_{0L}^{fip}|^2-|g_{0R}^{fip}|^2\right)
\int_0^1 {x^3dx\over{(1-x)m_{\chi^0_p}^2+x m_{{\tilde f}_i}^2}} \nonumber \\
{F}_{A,f}^{(a2)}(q^2)&=&-Q_{f^{\prime}} {\alpha M_Z^2\over{48\pi }}\sum_{i,p}
\left(|g_{L}^{fip}|^2-|g_{R}^{fip}|^2\right) 
\int_0^1 {x^3dx\over{(1-x)m_{\chi^+_p}^2+x m_{{\tilde f^{\prime}}_j}^2}} 
\nonumber \\
{F}_{A,f}^{(b)}(q^2)&=&2I_{3}^f{\alpha M_Z^2\over{48\pi }}\sum_{i,p}\left( |g_L^{fip}|^2- |g_R^{fip}|^2\right )\int_0^1
{x^2(x-3)dx\over{(1-x)m_{{\tilde f}_i}^2+x m_{\chi^+_p}^2}},
\eeqa
where $f^{\prime}$ stands for the isopartner of the 
fermion $f$ ({\it e.g.} if $f$ is the neutrino, then $f^{\prime}$ 
is the electron).   Notice that for quark anapole moment, an additional 
gluino contribution should be added, when 
parity is broken in the presence of a non-zero left-right mixing in the 
squark sector and non-equal diagonal left  and right squark masses. 
The gluino contribution can be 
obtained from ${F}_{A,f}^{(a1)}$ 
using the substitution rule give in Eq.~(\ref{eq:gluinosub}).
It should be remembered that the anapole moment of an elementary 
particle is not a physical observable \cite{michael}.
Here, we separate it from the other one-loop contributions purely 
for clarity of presentation.

\subsection{Box Graphs}

Let us introduce the following notation:
\beqa
L(f^\prime,f)_\lambda&=&{\bar f^\prime}(p^\prime)
\gamma_\lambda(1-\gamma_5)f(p) 
\nonumber \\
R(f^\prime,f)_\lambda&=&{\bar f^\prime}(p^\prime)\gamma_\lambda
(1+\gamma_5)f(p) ,
\eeqa
where $p$ and $p^\prime$ are the momenta of the incoming 
particle $f$ and outgoing $f^{\prime}$, respectively.

The four-point integration functions are defined as:
\beqa
\label{eq:cc-quark-boxd-expr}
B_1(M_1,M_2,m_1,m_2)&=&{\int_0^1dx}{\int_0^1dy}{\int_0^1dz}{{z(1-z)}
\over D_4(M_1,M_2,m_1,m_2)} \nonumber \\
B_2(M_1,M_2,m_1,m_2)&=&{\int_0^1dx}{\int_0^1dy}{\int_0^1dz}{{z(1-z)}
\over D_4^2(M_1,M_2,m_1,m_2)} \nonumber \\
D_4(M_1,M_2,m_1,m_2)&=&z[(1-x)M_1^2+xM_2^2] +(1-z)[(ym_1^2+(1-y)m_2^2] .
\eeqa
The explicit formulae for $B_1$ and $B_2$ are
\beqa
\label{eq:box-integrals}
&&B_1(M_1,M_2,m_1,m_2)={m_1^4\ln{m_1^2\over M_2^2}
\over{2(m_1^2-M_1^2)(m_1^2-m_2^2)(m_1^2-M_2^2)}} \nonumber \\
&&+{m_2^4\ln{m_2^2\over M_2^2}
\over{2(m_2^2-m_1^2)(m_2^2-M_1^2)(m_2^2-M_2^2)}} +
{M_1^4\ln{M_1^2\over M_2^2}
\over{2(M_1^2-m_1^2)(M_1^2-m_2^2)(M_1^2-M_2^2)}} \nonumber \\
&&B_2(M_1,M_2,m_1,m_2)={m_1^2\ln{M_2^2\over m_1^2}
\over{(m_1^2-M_1^2)(m_1^2-m_2^2)(m_1^2-M_2^2)}} \nonumber \\
&&+{m_2^2\ln{M_2^2\over m_2^2}
\over{(m_2^2-m_1^2)(m_2^2-M_1^2)(m_2^2-M_2^2)}} +
{M_1^2\ln{M_2^2\over M_1^2}
\over{(M_1^2-m_1^2)(M_1^2-m_2^2)(M_1^2-M_2^2)}} .
\eeqa

\subsubsection{Charged Current Box}
We define the matrix element of the box diagram as 
\beq
{\cal{M}}_{box}^{CC}=-i \frac{G_\mu}{\sqrt{2}} \delta_{B}^{CC} 
L(\mu,\nu_{\mu})_\lambda\times L(u,d)^\lambda +\ldots
\eeq
to factor out the overall Fermi-constant.  The dots denote 
other Dirac structures appearing 
in the amplitude that make negligible 
contributions (suppressed by $m_{\mu}$) to the total cross-section.

\begin{figure}
\resizebox{12. cm}{!}{\includegraphics*[40,390][440,640]{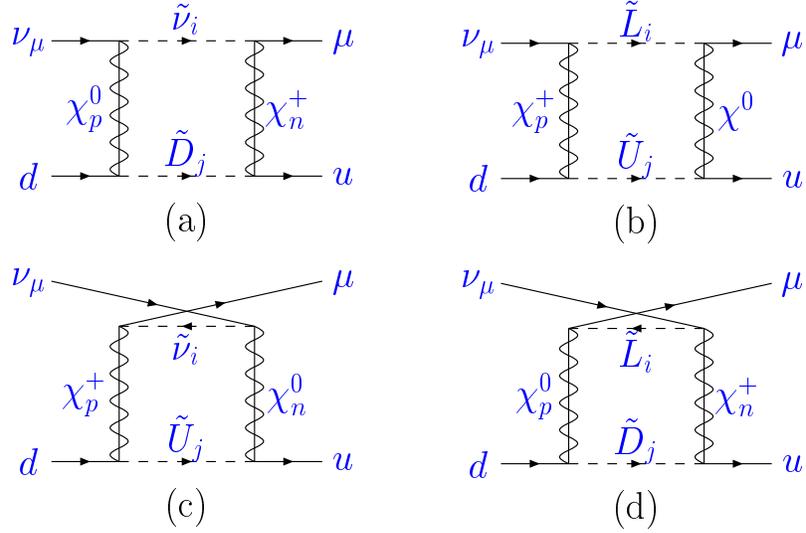}}
\caption{Feynman diagrams of one loop SUSY contribution to
neutrino charge current box diagram.}
\label{fig:box_CC}
\end{figure}

For the graphs in Fig. \ref{fig:box_CC} we have:
$\delta_B^{CC}=\delta_B^{CC;a}+\delta_B^{CC;b}+\delta_B^{CC;c}+\delta_B^{CC;d}$
\beqa
\label{eq:cc-box-expr}
\delta_{B}^{CC;a}&=&-{\alpha M_W^2 s^2 \over {4 \pi }}
{\sum_{p,n,i,j}}
g_{0L}^{\nu_\mu ip}g_{L}^{*\mu in}g_{0L}^{djp}g_{L}^{*ujn}
m_{\chi^0_p} m_{\chi^+_n}
B_2(m_{{\tilde \nu}_i},m_{{\tilde D}_j},m_{\chi^0_p},m_{\chi^+_n}) 
\nonumber \\
\delta_{B}^{CC;b}&=&-{\alpha M_W^2 s^2 \over {4 \pi}}
{\sum_{p,n,i,j}}
g_{L}^{{\nu_\mu}ip}g_{0L}^{*\mu in}g_{L}^{djp}g_{0L}^{*ujn} 
m_{\chi^+_p}m_{\chi^0_n} 
B_2(m_{{\tilde L}_i},m_{{\tilde U}_j},m_{\chi^+_p},m_{\chi^0_n}) 
\nonumber \\
\delta_{B}^{CC;c}&=&{\alpha M_W^2 s^2 \over {4 \pi}}
{\sum_{p,n,i,j}}
g_{0L}^{{\nu_\mu}in} g_{L}^{*\mu ip}g_{L}^{djp}g_{0L}^{*ujn} 
B_1(m_{{\tilde \nu}_i},m_{{\tilde U}_j},m_{\chi^+_p},m_{\chi^0_n}) 
\nonumber \\
\delta_{B}^{CC;d}&=&{\alpha M_W^2 s^2 \over {4 \pi }}
{\sum_{p,n,i,j}} 
g_L^{{\nu_\mu}in} g_{0L}^{*\mu ip} g_{0L}^{djp}g_L^{*ujn}
B_1(m_{{\tilde L}_i},m_{{\tilde D}_j},m_{\chi^0_p},m_{\chi^+_n}) .
\eeqa
Similarly, the box diagram contributing to the muon decay 
$\mu\rightarrow \nu_\mu e \bar\nu_e$ can be obtained from 
$\delta_B^{CC}$ using the substitution:
\beqa
&&\delta_B^{CC}\rightarrow \delta_B^{\mu *},\ \ 
d\rightarrow e,\ \ u\rightarrow \nu_e,\ \ 
\tilde{D}_j\rightarrow \tilde{L}_j,\ \ \tilde{U}_j \rightarrow 
\tilde{\nu}_j,\ \ 
\nonumber \\
&&g_{L,R,0L,0R}^{djp} \rightarrow g_{L,R,0L,0R}^{ejp},\ \ 
g_{L,R,0L,0R}^{ujn} \rightarrow g_{L,R,0L,0R}^{\nu_e jn} .
\eeqa

\subsubsection{Neutral Current Box}

We define the matrix element of the neutral current box diagram as 
\beq
{\cal{M}}_{box}^{NC}=-i \frac{G_\mu}{\sqrt{2}} 
\left[\delta_{B}^{L;q} L(\mu,\nu_{\mu})_\lambda\times L(q,q)^\lambda 
+\delta_{B}^{R;q}
L(\mu,\nu_{\mu})_\lambda\times R(q,q)^\lambda\right].
\eeq
The explicit expressions for the $\nu_\mu-q$ box graphs are given below.

\begin{figure}
\resizebox{16. cm}{!}{\includegraphics*[10,510][580,650]{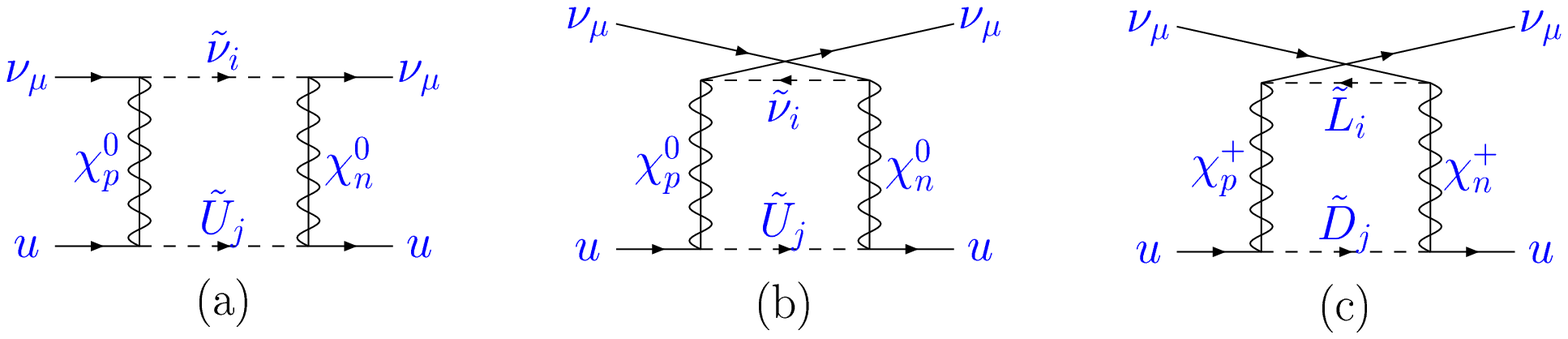}}
\caption{Feynman diagrams of one loop SUSY contribution to
$\nu_\mu-u$ neutrino neutral current box diagram.}
\label{fig:box_NC_u}
\end{figure}

\begin{figure}
\resizebox{16. cm}{!}{\includegraphics*[10,510][580,650]{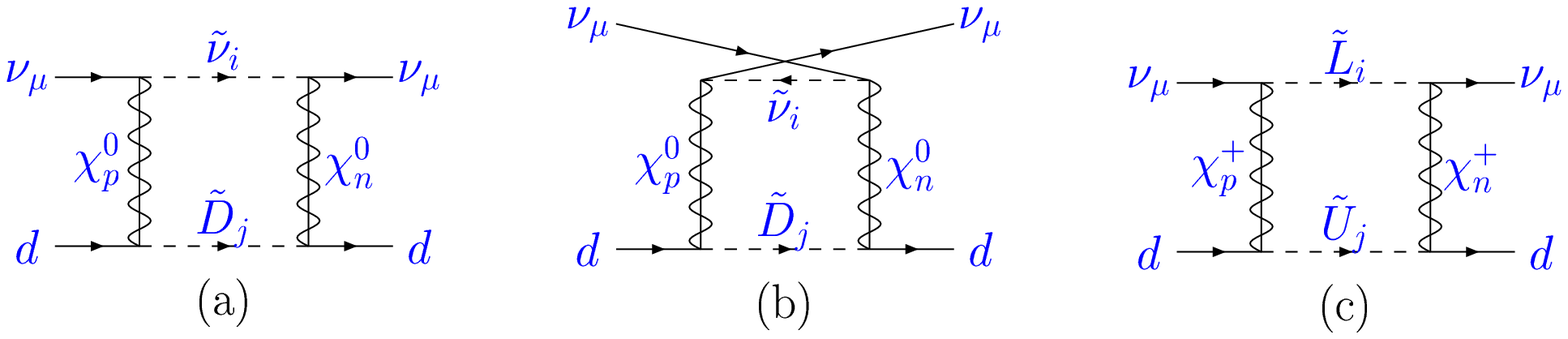}}
\caption{Feynman diagrams of one loop SUSY contribution to
$\nu_\mu-d$ neutrino neutral current box diagram.}
\label{fig:box_NC_d}
\end{figure}

Up-quark box diagrams (Fig.~\ref{fig:box_NC_u}):
$\delta_B^{L;u}=\delta_B^{L;(u,a)}+\delta_B^{L;(u,b)}+\delta_B^{L;(u,c)}$,
$\delta_B^{R;u}=\delta_B^{R;(u,a)}+\delta_B^{R;(u,b)}$,
\beqa
\label{eq:nc-u-box-expr}
\delta_{B}^{L;(u, a)}(NC)&=&-{\alpha M_W^2 s^2 \over {4 \pi }}
{\sum_{p,n,i,j}} 
g_{0L}^{\nu_\mu ip}g_{0L}^{*\nu_\mu in}g_{0L}^{ujp}g_{0L}^{*ujn}
m_{\chi^0_p} m_{\chi^0_n}
B_2(m_{{\tilde \nu}_i},m_{{\tilde U}_j},m_{\chi^0_p},m_{\chi^0_n})
\nonumber \\
\delta_{B}^{R;(u, a)}(NC)&=&-{\alpha M_W^2 s^2 \over {4 \pi }}
{\sum_{p,n,i,j}}
g_{0L}^{\nu_\mu ip}g_{0L}^{*\nu_\mu in}g_{0R}^{ujp}g_{0R}^{*ujn}
B_1(m_{{\tilde \nu}_i},m_{{\tilde U}_j},m_{\chi^0_p},m_{\chi^0_n})
\nonumber \\
\delta_{B}^{L;(u, b)}(NC)&=&{\alpha M_W^2 s^2 \over {4 \pi }}
{\sum_{p,n,i,j}}
g_{0L}^{\nu_\mu in}g_{0L}^{*\nu_\mu ip} g_{0L}^{ujp}g_{0L}^{*ujn}
B_1(m_{{\tilde \nu}_i},m_{{\tilde U}_j},m_{\chi^0_p},m_{\chi^0_n})
\nonumber \\
\delta_{B}^{R;(u, b)}(NC)&=&{\alpha M_W^2 s^2 \over {4 \pi }}
{\sum_{p,n,i,j}} 
g_{0L}^{\nu_\mu in}g_{0L}^{*\nu_\mu ip} g_{0R}^{ujp}g_{0R}^{*ujn}
m_{\chi^0_p} m_{\chi^0_n}
B_2(m_{{\tilde \nu}_i},m_{{\tilde U}_j},m_{\chi^0_p},m_{\chi^0_n})  
\nonumber \\
\delta_{B}^{L;(u, c)}(NC)&=&{\alpha M_W^2 s^2 \over {4 \pi}}
{\sum_{p,n,i,j}}
g_{L}^{\nu_\mu in}g_{L}^{*\nu_\mu ip}g_{L}^{ujp}g_{L}^{*ujn}
B_1(m_{{\tilde L}_i},m_{{\tilde D}_j},m_{\chi^+_p},m_{\chi^+_n}) .
\eeqa

Down-quark box diagrams (Fig.~\ref{fig:box_NC_u}):
$\delta_B^{L;d}=\delta_B^{L;(d,a)}+\delta_B^{L;(d,b)}+\delta_B^{L;(d,c)}$,
$\delta_B^{R;d}=\delta_B^{R;(d,a)}+\delta_B^{R;(d,b)}$,
\beqa
\label{eq:nc-d-box-expr}
\delta_{B}^{L;(d,a)}(NC)&=&-{\alpha M_W^2 s^2 \over {4 \pi}}
{\sum_{p,n,i,j}} 
g_{0L}^{\nu_\mu ip}g_{0L}^{*\nu_\mu in} g_{0L}^{djp}g_{0L}^{*djn}
m_{\chi^0_p} m_{\chi^0_n}
B_2(m_{{\tilde \nu}_i},m_{{\tilde D}_j},m_{\chi^0_p},m_{\chi^0_n}) 
\nonumber \\
\delta_{B}^{R;(d,a)}(NC)&=&-{\alpha M_W^2 s^2 \over {4 \pi}}
{\sum_{p,n,i,j}} 
g_{0L}^{\nu_\mu ip}g_{0L}^{*\nu_\mu in} g_{0R}^{djp}g_{0R}^{*djn}
B_1(m_{{\tilde \nu}_i},m_{{\tilde D}_j},m_{\chi^0_p},m_{\chi^0_n})
\nonumber \\
\delta_{B}^{L;(d,b)}(NC)&=&{\alpha M_W^2 s^2 \over {4 \pi}}
{\sum_{p,n,i,j}} 
g_{0L}^{\nu_\mu in}g_{0L}^{*\nu_\mu ip}g_{0L}^{djp}g_{0L}^{*djn}
B_1(m_{{\tilde \nu}_i},m_{{\tilde D}_j},m_{\chi^0_p},m_{\chi^0_n})
\nonumber \\
\delta_{B}^{R;(d,b)}(NC)&=&{\alpha M_W^2 s^2 \over {4 \pi}}
{\sum_{p,n,i,j}} 
g_{0L}^{\nu_\mu in}g_{0L}^{*\nu_\mu ip}g_{0R}^{djp}g_{0R}^{*djn}
m_{\chi^0_p} m_{\chi^0_n}
B_2(m_{{\tilde \nu}_i},m_{{\tilde D}_j},m_{\chi^0_p},m_{\chi^0_n}) 
\nonumber \\
\delta_{B}^{L;(d,c)}(NC)&=&-{\alpha M_W^2 s^2 \over {4 \pi}}
{\sum_{p,n,i,j}}
g_{L}^{\nu_\mu ip}g_{L}^{*\nu_\mu in}g_{L}^{djp}g_{L}^{*djn} 
m_{\chi^+_p}m_{\chi^+_n} 
B_2(m_{{\tilde L}_i},m_{{\tilde U}_j},m_{\chi^+_p},m_{\chi^+_n}) .\ \ \ 
\eeqa

\section{radiative correction to neutrino-nucleus interactions}
Given the expressions for the fermion field strength renormalization
$\delta Z_{L,R}^{F_I}$, vertex correction $\delta_{V F_I F_J}^{L,R}$ 
and box diagrams $\delta_B^{L,R}$, $\delta_{VB}^{CC}$ in 
Eq.~(\ref{eq:rhoCC}), $\delta_V^\nu$ in Eq.~(\ref{eq:rhoNC}) and 
$\delta_{VB}^{L,R;q}$ in Eq.~(\ref{eq:lambdaNC}) can be expressed as
\beqa
\delta_{VB}^{\mu}&=&\frac{1}{2}\left(
\delta Z_L^\mu + \delta Z_L^{\nu_\mu} + \delta Z_L^e + \delta Z_L^{\nu_e}
\right)- \sqrt{2}s \left(\delta_{W\nu_e e}^L+\delta_{W\mu\nu_\mu}^L \right)
+\delta_B^\mu\\
\delta_{VB}^{CC}&=&\frac{1}{2}
\left(\delta Z_L^{\nu_\mu}+\delta Z_L^{\mu}
+\delta Z_L^{u}+\delta Z_L^{d}\right)
-\sqrt{2}s \left(\delta_{W\nu_\mu \mu}^L + \delta_{W d u}^L\right)
+\delta_B^{CC}\\
\delta_{V}^{\nu}&=&\delta Z_L^{\nu_\mu}-2sc\ \delta_{Z\nu_\mu \nu_\mu}^L\\
\delta_{VB}^{L;q=u,d}&=&\frac{1}{2}\delta Z_L^{q}
-sc\ \delta_{Z q q}^L+\delta_B^{L;q}\\
\delta_{VB}^{R;q=u,d}&=&\frac{1}{2}\delta Z_R^{q}
-sc\ \delta_{Z q q}^R+\delta_B^{R;q} .
\eeqa

The neutrino anapole moment contribution to the neutral current 
neutrino-nucleus scattering is shown in Fig. \ref{fig:NC}(b), 
which is absorbed into $\delta\kappa_{\nu}$ as 
\beq
\delta\kappa=-4c^2 {F}_{A,\nu}(q^2).
\eeq


\end{document}